\documentclass[a4paper,fleqn,usenatbib,useAMS]{mnras}

\pdfoutput=1

\usepackage{graphicx}	% Including figure files
\usepackage{amsmath}	% Advanced maths commands
\usepackage{amssymb}	% Extra maths symbols
\usepackage{multicol}        % Multi-column entries in tables
\usepackage{bm}		% Bold maths symbols, including upright Greek
\usepackage{pdflscape}	% Landscape pages
\usepackage[section]{placeins}

\usepackage{aas_macros}
\usepackage{natbib}
\usepackage{hyperref}
\usepackage{xcolor}
\usepackage{pdfcomment}
\usepackage{soul}
\usepackage{float}
\usepackage{subfig}
\usepackage{rotating}

\usepackage[T1]{fontenc}
\usepackage{ae,aecompl}

\usepackage{newtxtext,newtxmath}
\title[The nuclear cluster of NGC\,6384 with LBT/ARGOS/LUCI\,2]
{
The Milky Way like galaxy NGC\,6384 and its nuclear star cluster at high NIR spatial resolution using LBT/ARGOS commissioning data
}

\author[I. Y. Georgiev et al.]
{
Iskren Y. Georgiev$^1$\thanks{e-mail: \href{mailto:georgiev@mpia.de}{georgiev@mpia.de}
}, %\thanks{Present address: Science magazine, AAAS Science International, \mbox{82-88}~Hills Road, Cambridge CB2~1LQ, UK}
Nadine Neumayer$^1$, Wolfgang G\"assler$^1$, Sebastian Rabien$^2$, 
\and\hspace{.1cm}Lorenzo Busoni$^3$, Marco Bonaglia$^3$, Julian Ziegleder$^2$, Gilles Orban de Xivry$^5$, Diethard
\and\hspace{.1cm}Peter$^1$, Martin Kulas$^1$, Jose Borelli$^1$, Gustavo Rahmer$^4$, Michael Lefebvre$^4$, Holger Baumgardt$^6$\\
$^1$Max-Planck Institut f\"ur Astronomie, K\"onigstuhl 17, 69117 Heidelberg\\
$^2$Max-Planck Institut f\"ur Extraterrestrische Physik, Giessenbachstrasse 1, 85748 Garching, Germany\\
$^3$Arcetri Astrophysical Observatory, Largo Enrico Fermi 5, 50125 Florence, Italy\\
$^4$LBT Observatory, University of Arizona, 933 N. Cherry Ave, Tucson, AZ 85721, USA\\
$^5$Universit\'e de Li\`ege, Institut d'Astrophysique et de G\'eophysique (B\^at. B5c), Quartier Agora, All\'ee du 6 ao\^ut, 19C, B-4000 Li\`ege 1 (Sart-Tilman), Belgique \\
$^6$University of Queensland, St. Lucia, QLD 4068, Australia
}

\date{Last updated yyyy/mm/dd; in original form yyyy/mm/dd}

\pubyear{2018}

\begin{document}
\label{firstpage}
\pagerange{\pageref{firstpage}--\pageref{lastpage}}
\maketitle

\begin{abstract}
We analyse high spatial resolution near infra-red (NIR) imaging of NGC\,6384, a Milky Way like galaxy, using 
ARGOS commissioning data at the Large Binocular Telescope (LBT). ARGOS provides a 
stable PSF$_{\rm FWHM}\!=\!0\farcs2\!-\!0\farcs3$ AO correction of the ground layer 
across the LUCI\,2 NIR camera $4\arcmin\!\times4\arcmin$ field by using six laser guide 
stars (three per telescope) and a na\-tural guide star for tip-tilt sensing and guiding. 
Enabled by this high spatial resolution we analyse the structure of the nuclear star 
cluster (NSC) and the central kiloparsec of NGC\,6384. We find via 2D mo\-dell\-ing that 
the NSC ($r_{\rm eff}\!\simeq\!10$\,pc) is surrounded by a small 
($r_{\rm eff}\!\simeq\!100$\,pc) and a larger Sersi\'c ($r_{\rm eff}\!\simeq\!400$\,pc), 
all embedded within the NGC\,6384 large-scale boxy/X-shaped bulge and disk. This 
proof-of-concept study shows that with the high spatial resolution achieved by 
ground-layer AO we can push such analysis to distances previously only accessible from 
space. SED-fitting to the NIR and optical HST photometry allowed to leverage the 
age-metallicity-extinction degeneracies and derive the effective NSC properties of 
an young to old population mass ratio of $8\%$ with ${\cal M}_{\rm\star,old}\!\simeq\!3.5\times10^7M_\odot$, 
Age$_{\rm old,\ young}\!=\!10.9\pm1.3$\,Gyr and 226\,Myr $\pm62\%$, metallicity 
[M/H]$=\!-0.11\pm0.16$ and $0.33\pm39\%$\,dex, and $E(B\!-\!V)\!=\!0.63$ and 1.44\,mag.
\end{abstract}


%%%%%%%%%%%%%%%%%%%%%%%%%%%%%%%%%%%%%%%%%%%%%%%%%%

%%%%%%%%%%%%%%%%% BODY OF PAPER %%%%%%%%%%%%%%%%%%

% The MNRAS class isn't designed to include a table of contents, but for this document one is useful.
% I therefore have to do some kludging to make it work without masses of blank space.
%\begingroup
%\let\clearpage\relax
%\tableofcontents
%\endgroup
%\newpage

\section{Introduction}

\begin{figure}
\includegraphics[width=.463\textwidth]{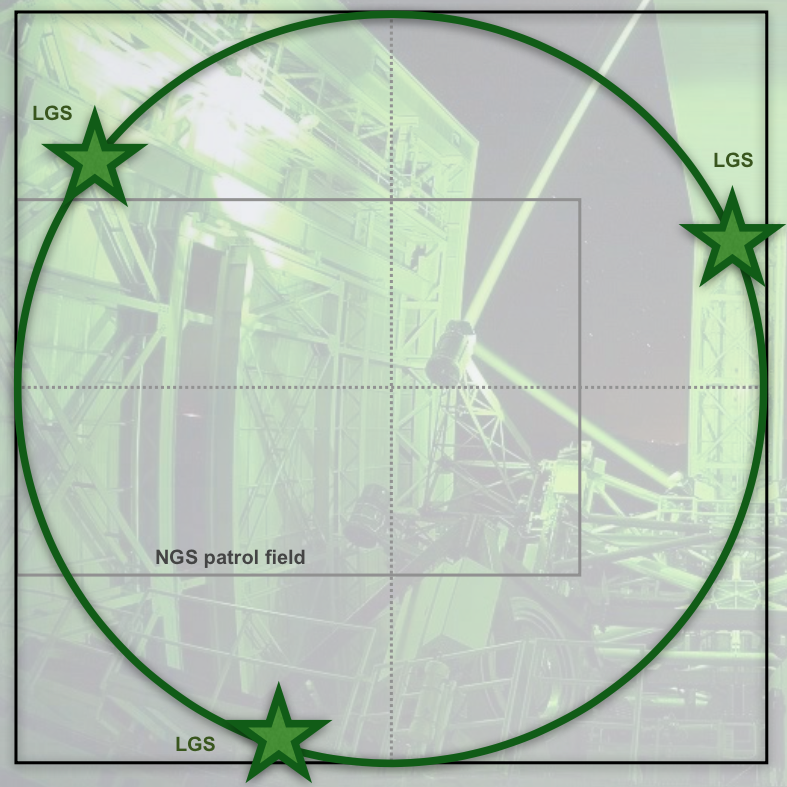}
\caption{Schematic view of the $4\arcmin\times4\arcmin$ LUCI\,2 field of view (solid, 
black outer square), the $\sim3\arcmin\times2\arcmin$ field of the NGS camera board 
(light, gray rectangle), and the $2\arcmin$ radius circle (thick, green line circle), 
where the three LGSs are located within the LUCI FoV.
}\label{fig:ARGOS_LUCI}
\end{figure}

The development of current telescope technology is \emph{essential} to increase the 
efficiency of scientific output. In particular, high spatial resolution over a large field 
of view is one such domain that was recently commissioned at the Large Binocular Telescope 
\cite[LBT\footnote{\href{http://www.lbto.org/}{http://www.lbto.org/}},][]{Hill94,Hill12} 
on Mount Graham in Arizona. The Advanced Rayleigh guided Ground layer adaptive Optics System, 
ARGOS\footnote{\href{http://www.mpe.mpg.de/ir/argos}{http://www.mpe.mpg.de/ir/argos}} \citep{Rabien18}, 
equips the LBT with six green light (532\,nm) lasers (three on each side situated on a circle 
with 2\arcmin\ radius (see Fig.\,\ref{fig:ARGOS_LUCI}). 
ARGOS corrects the ground layer turbulence by using artificial laser stars focused at 12\,km 
and a tip-tilt star for wave front sensing and guiding. This improves the near infra-red 
(NIR) natural seeing by a factor of 2 -- 3, and provides a stable Point Spread Function 
(PSF) over a $4\arcmin\times4\arcmin$ field of view with $\lesssim\!20\%$ increase toward the 
detector edges. Together with the NIR cameras LUCI\,1\,\&\,2 (see \S\,\ref{Sect:DataReduction}) 
this makes LBT uniquely equipped for high-spatial resolution NIR science. A dedicated ARGOS 
paper describing its technical aspects, performance and commissioning results is published 
by \cite{Rabien18}. The aim of this paper is to utilize the large field of view AO correction 
and push the limits to study the structure of compact star clusters beyond distances accessible 
for ground based seeing limited observations.

Here we focus on the spatial analysis of the nuclear star cluster (NSC) and its host 
galaxy NGC\,6384 at a distance of 20.7\,Mpc \cite[$m\!-\!M\!=\!31.58$\,mag, ][]{Sorce14}. 
In a forthcoming paper we will present the analysis of the star cluster population of NGC\,6384. 
This galaxy was selected as a first scientific commissioning target due to the large amount 
of Galactic stars suitable for performance testing. In addition, its stellar mass 
\cite[$\sim5\times10^{10}M_\odot$, ][]{Georgiev16} and boxy/peanut bulge structure 
\citep{Erwin&Debattista13} makes it appear very similar to the Milky Way (MW). Its NSC 
is more extended but of comparable mass \cite[$r_{\rm eff,V}\!\sim\!15$\,pc, ${\cal M}\!\simeq\!1.3\times10^7M_\odot$,][and found here in the NIR]{Georgiev&Boeker14,Georgiev16} 
to that of the MW nuclear cluster \cite[$r_{\rm eff,K_S}\!=\!4.2$\,pc, ${\cal M}\!=\!2.5\times10^7M_\odot$,][]{Schoedel14}. 
The similar mass might suggest that it is possible to host a similarly massive black hole (MBH) of $4.1\times10^6M_\odot$ 
\citep{Gravity_MW_MBH_18,Feldmeier17}. A potential MBH would also explain its 
classification as a LINER galaxy \citep{Ho97}. This makes it a good test case in studying 
the NSC NIR properties of a MW analogue at $\sim\,21$\,Mpc. The main difference to the 
MW, however, is its very low galaxy density environment. To within $\pm400$\,km/s of its 
systemic velocity (1664\,km/s)\footnote{Velocity information has been retrieved from the 
\href{http://leda.univ-lyon1.fr/}{HyperLeda} database \citep{Makarov14}.} the nearest bright 
galaxy is NGC\,6509 (1813\,km/s, $\sim2.5$\,mag fainter than NGC\,6384) at a projected 
separation of $\sim2.5$\,Mpc ($\sim6.7^\circ$) and is also twice less massive 
\cite[$\sim2\times10^{10}M_\odot$, ][]{Georgiev16}.

Due to the high density nature of NSCs\,\cite[compact\,$\sim\!4$\,pc and massive, 
$\sim\!10^6M_\odot$, e.g.][]{Boeker02,Boeker04,Walcher05,Georgiev&Boeker14,Georgiev16,Ordenes-Briceno18,Sanchez-Janssen18}, 
obtaining high spatial resolution data in the NIR is crucial to break the age-metallicity-reddening 
degeneracies present in the optical alone \cite[e.g.][]{deMeulenaer14}. This is very important for studying their properties, 
where young stellar populations and high extinction are expected to be present. Spatially 
resolving these provides important constraints on our understanding of how galactic nuclei 
and NSCs build up, e.g. via cluster migration and merging \cite[e.g.][]{Tremaine75,Capuzzo-Dolcetta93,Oh&Lin00,Bekki04,Antonini12,Tsatsi17} 
or \emph{in situ} from gas accretion \cite[e.g.][]{Boeker03,Milosavljevic04,Schinnerer03,Schinnerer06,Bekki07} 
or most likely a combination of both \cite[e.g.][]{Neumayer11,Hartmann11,Antonini15,Cole17,Kacharov18}.
In addition, studies of NSCs have shown that their size increases with wavelength 
\citep{Georgiev&Boeker14,Carson15}, which is consistent with earlier findings that NSCs 
contain morphologically distinct young and old stellar populations \citep{Seth08b}, where 
the young one is more centrally concentrated. From numerical simulations is known 
that due to energy equipartition and orbital relaxation merging of clusters typically 
leads to a final cluster with larger effective radius ($r_{\rm eff}$) 
\cite[e.g.][]{Fellhauer&Kroupa02a,Bekki04}. On the other hand NSC growth dominated by 
episodic gas accretion will lead to a more compact clusters \cite[e.g.][]{Tsang&Milosavljevic18}.

Optical-NIR Spectral Energy Distribution (SED) modelling of diffuse light \cite[e.g.][]{Carson15,Dale16} 
at high spatial resolution can provide a map of the spatial variation and composition 
of the main stellar population components of the NSC, as well as their host galaxy. 
Such analysis can also unveil (an additional) contribution to the optical-NIR light 
from accretion onto an obscured nuclear MBH and/or nuclear star formation activity 
\cite[e.g.][]{Noll09,Drouart16}. To understand NSC formation, it is therefore critical 
to be able to disentangle such degeneracies. For example, follow-up spectroscopic 
observations aiming at decomposing the main stellar populations and the dynamical 
imprint of a MBH rely on a good mass model to predict the stellar population velocity 
profile \cite[e.g.][]{Neumayer06,Seth10,Neumayer11,Neumayer&Walcher12,Nguyen18}. Observations of NSCs have 
shown that although they contain young stellar populations and extended star formation 
histories (SFHs), the most dominant one by mass is the oldest ($\gtrsim\!3$\,Gyr), where 
more than 50\% of the mass of the cluster has formed \cite[e.g.][from spectral modelling in the optical]{Walcher06,Kacharov18}. 
Therefore, characterising the spatial structure of NSCs in the NIR is of particular 
importance, because that is where most of the stellar light of the old stellar population 
is emitted that allows us to trace most of the gravitating mass. Therefore, 
characterizing NSCs in the NIR can provide additional constraints as to which of 
the leading NSC formation scenarios had a major role in the formation of particular NSC. 
However, to be able to achieve this for a larger sample and of more distant galaxies, 
efficient high spatial resolution observations are required, such as with the presented 
here wide-field ground-layer AO NIR observations in the NIR with ARGOS at the LBT.

Here we first present the observations, data reduction and calibrations in 
\S\,\ref{Sect:DataReduction}. In \S\,\ref{Sect:Analysis} we present our analysis of the 
data with the colour-magnitude diagram (CMD, in \S\,\ref{Sect:Phot}) of all sources in 
the LUCI\,2 field of view of NGC\,6384 and in \S\,\ref{Sect:Struct:NSCGal} we present 
the structural analysis of the NSC and NGC\,6384. In \S\,\ref{Sect:SED} we combine the 
optical photometry from archival HST data and the LUCI\,2 NIR photometry to derive via 
SED fitting the NSC effective age, luminosity weighted mass, metallicity and reddening. 
Finally, in \S\,\ref{Sect:Discussion} we discuss and summarize our findings.

\section{Observations, reduction and calibration}\label{Sect:DataReduction}

\subsection{Near infar-red LBT/LUCI\,2 data and PSF photometry}\label{Sect:DataReduction.LUCI2}

\begin{figure}
\includegraphics[width=0.48\textwidth]{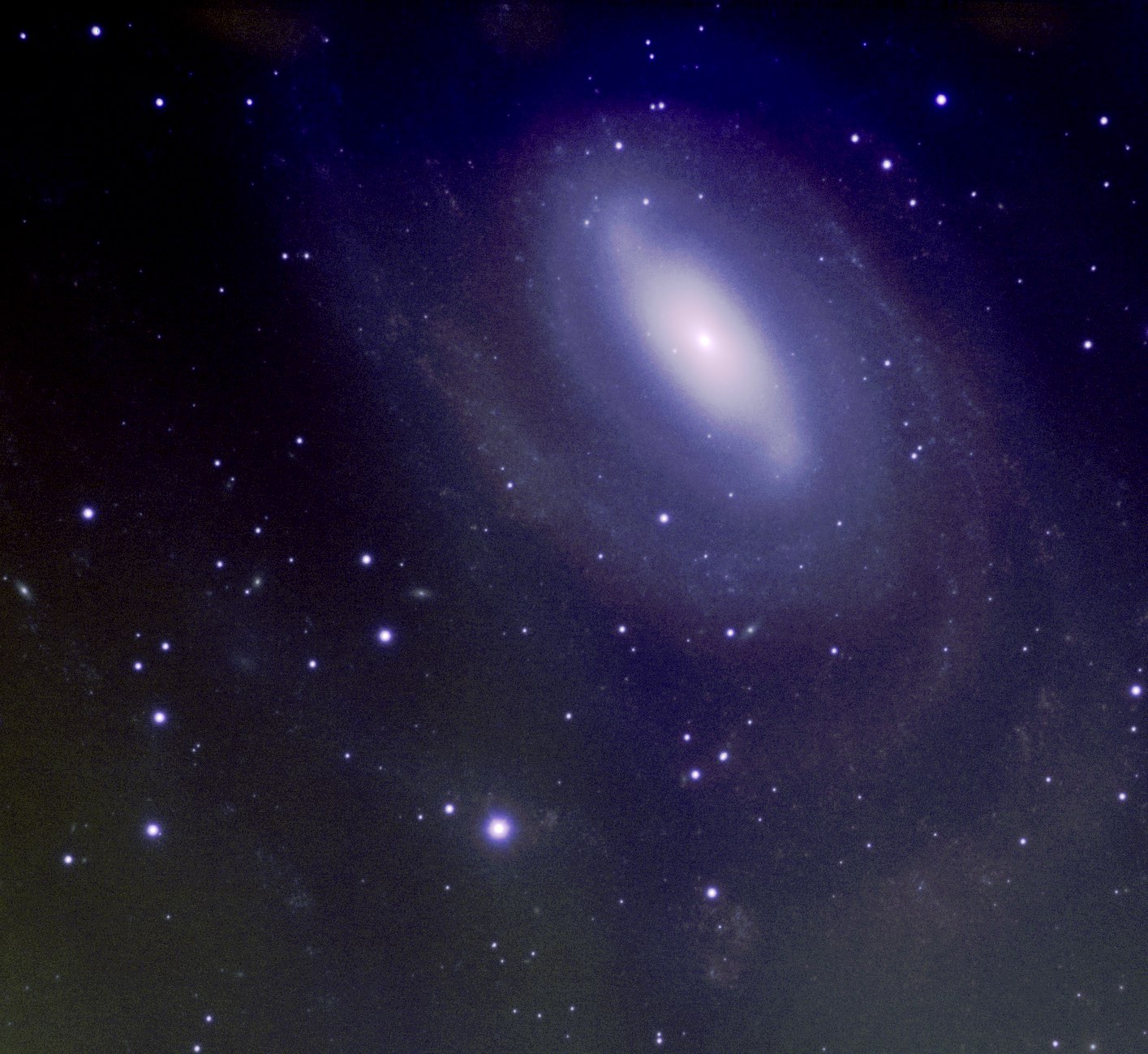}
\caption{$JHK_S$ colour composite image of a $2\farcm5\times2\farcm5$ ($\sim18\times18$\,kpc) 
region of NGC\,6384 taken with the LUCI\,2 camera on LBT during the first ARGOS 
commissioning run in 2015-05-02. The achieved spatial resolution across the 
LUCI\,2 4'\,x\,4' field is $\sim\!0.25\arcsec\pm20\%$. The intensity scale is stretched 
such that to show the compact nucleus, the bar and X-shaped boxy bulge and spiral arms 
of NGC\,6384. North is up, and East is to the left.
}\label{fig:N6384_JHKs_colour}
\end{figure}

NGC\,6384 imaging data was obtained during the ARGOS commissioning run on 2015-05-01 
and 02. The total integration time in closed loop with the LBT/LUCI\,2 NIR camera 
\citep{Seifert03} and a Detector Integration Time (DIT)=5\,s is $J\ (775s),H\ (860s),K_S\ (70s)$, i.e. 155, 172 
and 14 frames per filter, respectively. Due to constraints from commissioning tasks, 
the $K_S$-band data was not fully completed. Nevertheless, as we will show later 
(\S\,\ref{Sect:Phot}), the depth was sufficient enough for our analysis of the NSC 
and the brightest GCs. The latter will be investigated in a  forthcoming study, while 
here we focus our analysis only on the NSC and the central kiloparsec of NGC\,6384. 
A colour composite $JHK_S$ image of NGC\,6384 shown in Figure\,\ref{fig:N6384_JHKs_colour} 
is zoomed at the central $\sim\!3\arcmin\times3\arcmin$. The intensity scaling is 
adjusted such that it better shows various galactic structures (nucleus, bar, bulge, 
spiral arms). The brightest star (middle-bottom) is the used as the tip-tilt star (a Natural 
Guide Star, NGS, for Wave Front Sensing, WFS). The LUCI\,2 plate scale of $0\farcs1189/$\,pix 
corresponds to 12\,pc/pix at the distance to NGC\,6384 of 20.7\,Mpc. We used a rectangular 
dither pattern within $\sim12\arcsec$ to eliminate detector cosmetics (bad pixels). 
The dither pattern is also chosen such that the NGS stays within the field of view of 
the NGS camera board, shown with gray solid rectangle in Figure\,\ref{fig:ARGOS_LUCI}. 
In this schematic representation one can see that the ARGOS LGSs are situated on a circle 
with a radius of $2\arcmin$, where 
their exact location depends on the field orientation. Off-target sky frames beyond the 
galaxy extent were unfortunately not obtained due to technical limitations during this 
first science commissioning run. Nevertheless, we are able to obtain a good global 
photometric calibration by using about 30 stars in common with the 2\,MASS point source 
catalogue (details follow below).

Image reduction was performed with tasks in {\it IRAF}\footnote{{\it IRAF} is 
the Image Reduction and Analysis Facility, a general purpose software system for 
the reduction and analysis of astronomical data. IRAF is written and supported 
by the National Optical Astronomy Observatories (NOAO) in Tucson, Arizona. NOAO 
is operated by the Association of Universities for Research in Astronomy (AURA), 
Inc. under cooperative agreement with the National Science Foundation}. We first 
correct all data frames for pixel non-linearity and persistence (see \S\,\ref{Sect:App.NonLin}). 
Although LUCI\,2 is cryogenically cooled, we create a master dark frame to account 
for the low level thermal dark current. Flats taken at the end of the observation 
were used to obtain a normalized master flat to correct for pixel-to-pixel differences 
in quantum efficiency, illumination and sensitivity across the detector. 
We create a bad pixel map from the ratio between low and high count flats. 
From each science exposure we subtracted the master dark and divided by the 
normalized flat. The final image registration and coaddition was performed in 
identical manner as in \cite{Georgiev08,Georgiev12,Georgiev&Boeker14} with a self 
developed IRAF procedure\footnote{A wrapper routine for the IRAF procedures {\sc 
daofind, daophot, allstar, geomap, geotran, imcombine}.}. It matches stars across 
all exposures, computes geometric distortions, uses the bad pixel map to flag pixels 
and a $15\!\times\!15$ pixel image statistic region to compute zero level offsets 
with respect to a reference exposure. Using pixel statistic region guarantees 
minimum biases and introduction of noise to objects' fluxes due to eventual residuals 
from imperfect background sky subtraction. This way we obtained the final coadded 
images in each band. We note that the same physical location of the statistic region 
was adopted for all filters. This is important in order to minimize biases in the 
final photometric colours.

Thanks to the ARGOS ground layer atmospheric correction, our final combined images 
achieved an excellent resolution of $PSF_{\rm FWHM}\!\simeq\!0\farcs2$ from the 
central tip tilt star to $\leq\!0\farcs3$ at the detector borders (see details in 
\S\,\ref{Sect:App.PSF}). To properly account for the variable PSF across the 
detector, we used the large number of Galactic foreground stars ($>\!100$) to 
build a spatially variable PSF model in each filter. The model\footnote{The empirical 
PSF was best fit by a single Moffat function, unlike the complex PSF of full AO systems.} 
is used to perform a PSF photometry on all detections above $\geq\!2\times\sigma_{\rm bckgr}$ threshold 
with a PSF model radius of 9\,pix\,$\simeq1\arcsec\simeq4\times PSF_{\rm FWHM}$. 
The aperture correction is derived from curve of growth aperture photometry of 
the PSF stars.

The photometric zeropoints are obtained from the error weighted least squares 
fit through the difference between the 2MASS and instrumental magnitude against 
2MASS magnitude and colour that yield the photometric zeropoints in $J,\ H$ and 
$K_S$ of $25.891\ \pm0.013,\ 25.654\ \pm0.013$ and $24.922\ \pm0.023$\,mag, 
respectively. We also checked for $J\!-\!K_S$ colour dependence, and found it 
to be negligible compared to the photometric uncertainties, which is driven by 
the faintest stars with worst 2MASS photometry and most affected by the background 
noise. Finally, the magnitudes are corrected for foreground Galactic reddening 
$A_J,\ A_H\ {\rm and}\ A_{Ks}=0.087,\ 0.055\ {\rm and}\ 0.037$\,mag based on 
$E(B-V)\!=\!0.11$\,mag \cite{Schlafly11} recalibration of the \cite{Schlegel98} 
Galactic dust maps and assuming \cite{Fitzpatrick99} reddening law with $R_V=3.1$.

\subsection{Optical HST data reduction and PSF model}\label{Sect:DataReduction.HST}

\begin{figure*}
\includegraphics[width=0.45\linewidth, bb=85 20 335 240]{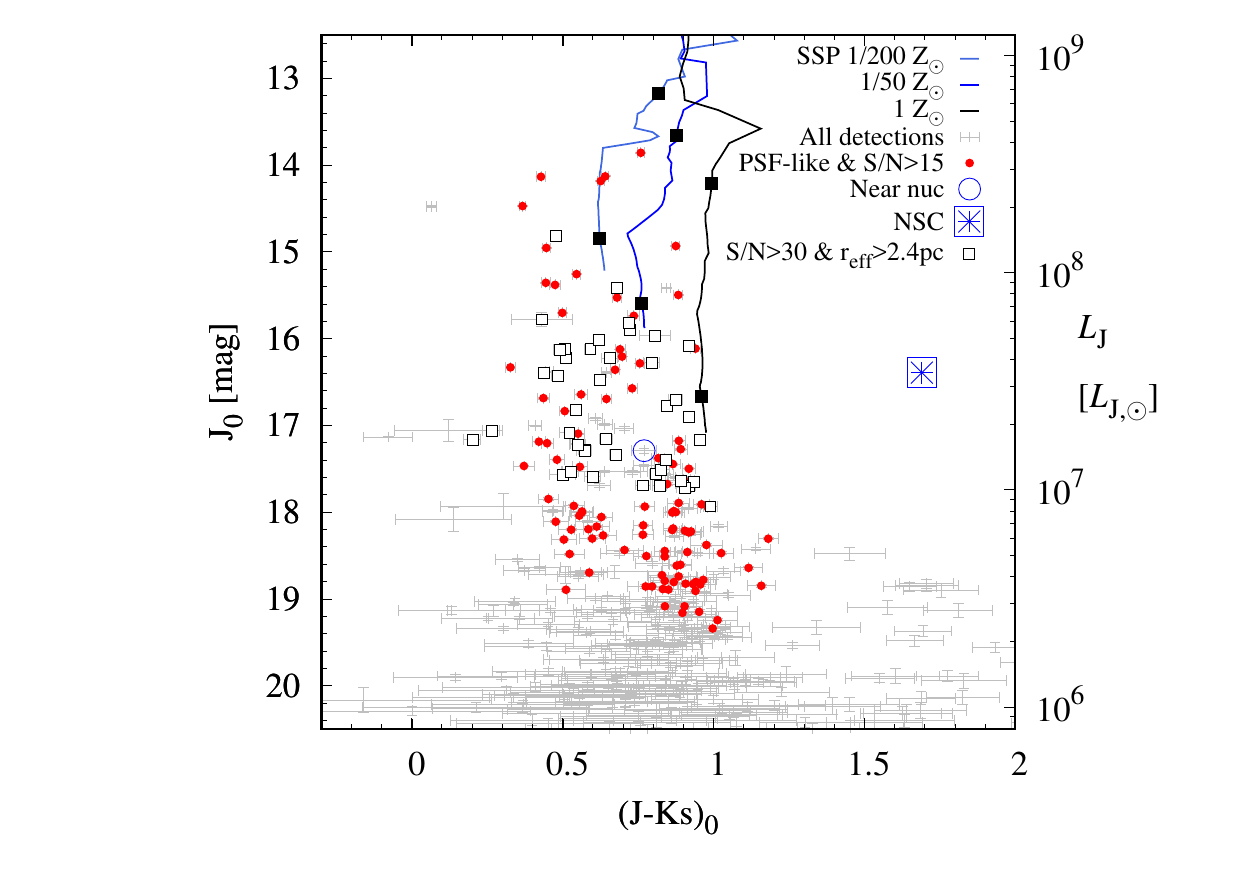}
\includegraphics[width=0.45\linewidth, bb=45 20 295 240]{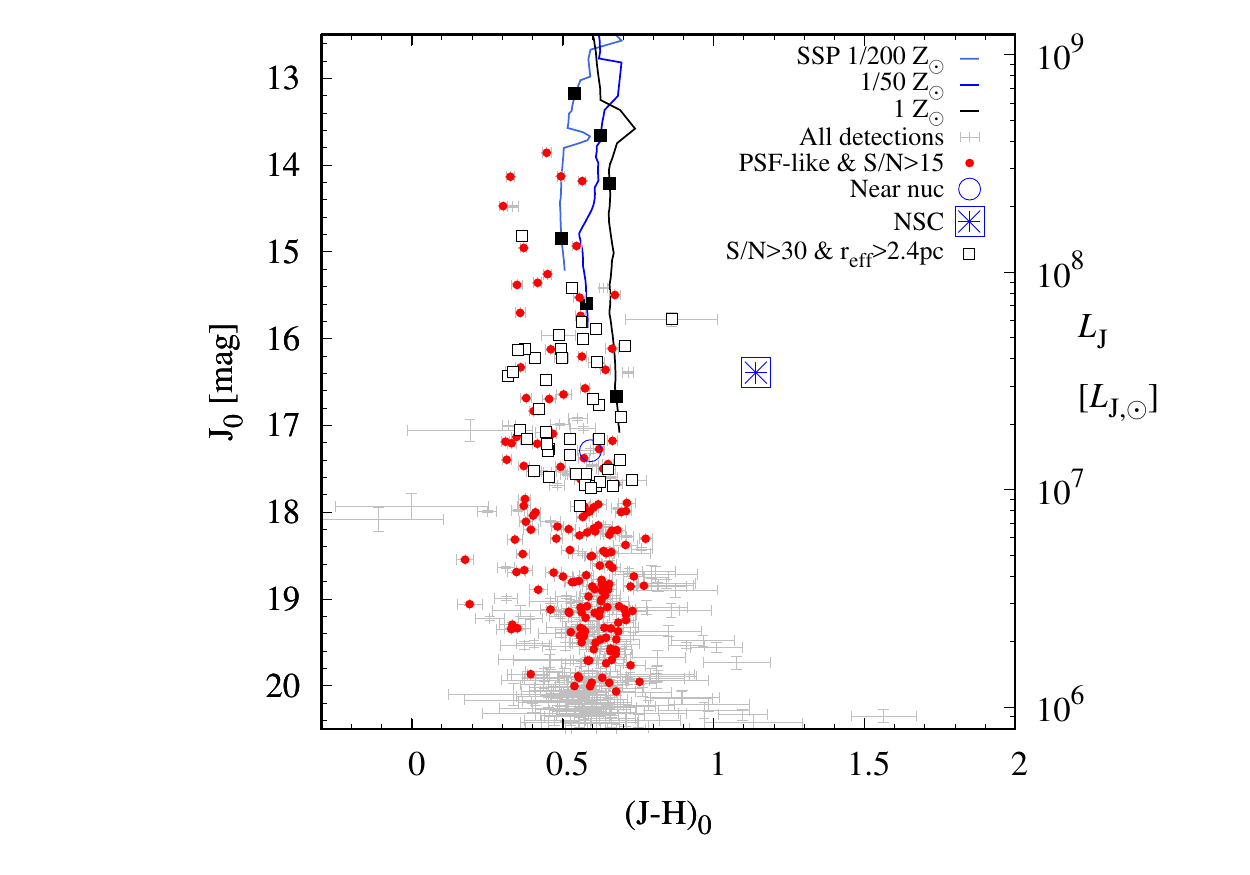}
\caption{Foreground Galactic deredenned colour-magnitude diagrams $J_0$\,vs.\,$(J\!-\!K_S)_0$ 
({\bf left}) and vs.\,$(J\!-\!H)_0$ ({\bf right}) of sources in the $4\arcmin\times4\arcmin$ 
LUCI\,2 field of NGC\,6384. Gray symbols with errorbars are all detections. Solid 
(red) circles are $S/N\!>15$ sources with a stellar-like PSF. Open squares are sources 
with $S/N\!>\!30$ for which the additional PSF analysis suggests a size bigger than 
10\% of the PSF$_{\rm FWHM}$, i.e. $>\!2.4$\,pc at the distance of NGC\,6384. 
The large open (blue) circle is a bright near nuclear source suspected to be a massive star cluster 
projected close to the NSC. Asterisk indicates the integrated NSC magnitude and colour. 
For comparison, with solid (nearly vertical) lines we show scaled \protect\cite{BC03} 
SSP models with metallicity of a solar, 1/50 ([Z/H$]\!=\!-0.7$\,dex) and 1/200 ([Z/H$]\!=\!-2.3$\,dex), 
as indicated in the figure legend. Two solid (black) squares along each SSP track 
indicate an age of 2 and 14\,Gyr from top to bottom, respectively. The model 
tracks are scaled such that the most metal-poor track at 14Gyr has roughly $10^8L_{\rm J_\odot}$.
\label{fig:CMD}
}
\end{figure*}

We retrieved from the HST archive WFPC\,2 ($F606W$) and ACS/WFC ($F475W,\ F814W$) 
data from Cycle\,6 and 11 programs SNAP-6359 (PI: M. Stiavelli) and GO-9395 (PI: 
M. Carollo), respectively. We used the latest version of the {\sc astrodrizzle} 
routine in PyRAF\footnote{\href{http://www.stsci.edu/institute/software_hardware/pyraf}{PyRAF} 
is a product of the Space Telescope Science Institute, which is operated by AURA 
for NASA. PyRAF is python environment for running IRAF tasks} to reprocess the 
archival data in order to achieve optimal drizzling pixel fraction and final pixel 
scale. Since the central $\sim5\times5$\,kpc of NGC\,6384 are fully covered by 
the higher resolution WFPC\,PC chip (0.05\arcsec/pix), we restricted our analysis 
to only this WFPC\,2 detector. Its resolution is nearly identical to that of the 
ACS WFC (0.04\arcsec/pix), therefore we chose to drizzle the WFPC2/PC1 and the ACS 
images to the same final pixel scale of 0.05\arcsec/pixel. We did not go for a smaller 
pixel scale due to the limited number of only two single exposures. We found that 
the optimal drizzle pixel fraction to be 0.85. Due to the large extent of the galaxy, 
we did not allow {\sc astrodrizzle} to automatically derive the sky value, instead, 
we provided our own. This value was obtained from the outermost galaxy sections 
on the ACS detectors. We obtained camera, detector, filter and 
position specific PSF model with the {\sc tinytim} software packge \citep{Krist&Hook11}, 
which we drizzled with the identical {\sc astrodrizzle} set up for the science images. 
Similarly to \S\,\ref{Sect:DataReduction.LUCI2}, the filter specific Galactic reddening 
toward NGC\,6384 is $A_{F435W}\!=\!0.434$\,mag, $A_{F606W}\!=\!0.293$\,mag, 
$A_{F814W}\!=\!0.180$\,mag.

\section{Analysis}\label{Sect:Analysis}

In this section we present the derived photometric properties of all sources in the 
LUCI\,2 field of NGC\,6384, including candidate star clusters whose detailed analysis 
we will present in a forth coming study. 
We derive the structure of the nuclear star cluster by fitting simultaneously its 
surroundings ($10\times10$\,kpc). We also analyse in identical manner the archival 
HST/ACS images. Using SED fitting of the NIR-optical best fit model magnitudes we 
derive luminosity weighted integrated properties of the NSC (e.g. stellar populations 
mass, age, metallicity, etc.). Similarly, we also compute a 1D projected of the surface 
mass density, metallicity, stellar population composition (young and old).

\subsection{Colour-magnitude diagram of the NGC\,6384 field}\label{Sect:Phot}

First, to illustrate the importance of the $K_S$-band in stellar population analysis 
in Figure\,\ref{fig:CMD} left and right we compare $J-K_S$ and $J-H$ colour-magnitude 
diagrams (CMDs), respectively, of all three filter matched sources with PSF photometry. 
Due to technical and time limitations during the ARGOS commissioning run, the 
$K_S-$\,band data did not reach as deep as in the $J-$ and $H-$bands. 
Although shallower, the $J\!-\!K_S$ colour index is scientifically more 
informative for the metallicity distribution of the sources, because it largely sensitive 
to the stellar effective temperature, which in turn is sensitive to metallicity. This 
can be appreciated by the comparison with Single Stellar Population (SSP) model 
tracks from \cite{BC03} shown with lines in Figure\,\ref{fig:CMD} for three metallicities 
indicated in the figure legend. Evidently, the dynamical range in the $J\!-\!K_S$ colour 
and separation between the SSP models of different metallicity is significantly larger 
than the colour uncertainty, which is important for estimating photometric metallicity.

Second, to demonstrate the achieved high spatial resolution over the entire 
$4\arcmin\times4\arcmin$ field with solid (red) circles we show sources with $S/N\!>\!15$ 
and stellar PSF, i.e. their sharp value\footnote{The sharp value is roughly the difference 
between object and stellar PSF returned by {\sc allstar} during the PSF photometry. Values 
close to 0 are stars, large positive bulges of background unresolved galaxies, and negative 
values are bad/hot pixels or cosmic rays.} 
is within |sharp|=0.15. Open squares indicate sources with $S/N\!\geq\!30$ and effective 
radius $r_{\rm eff}\!\geq\!2.4$\,pc as measured with the {\sc ishape} code \citep{Larsen99}. 
For high $S/N\!\geq\!30$ sources \cite{Larsen99} has shown that reliable size 
measurements can be performed down to $\sim10\%$ of the PSF$_{\rm FWHM}$. In our case 
this spatial ``resolution'' limit corresponds to $r_{\rm eff}\!\geq\!2.4$\,pc at the 
distance to NGC\,6384. The size measurement with {\sc ishape} was performed analogous to 
\cite{Georgiev08,Georgiev&Boeker14}. In brief, {\sc ishape} performs a $\chi^2$ 
minimization fitting of the object's profile with an analytical function convolved 
with a PSF model tailored to the objects' position on the detector (cf. \S\,\ref{Sect:DataReduction.HST}). 
We fitted \cite{King62} profiles with fixed and variable concentration index parameter 
$r_t/r_c=C=5,15,30,100$. The model with the best $\chi^2$ fit was adopted to be the 
model for the object's structural parameters. The analysis of the extended objects 
(some likely GC candidates) will be presented in a forthcoming study since our aim 
here is to focus on the NSC and NGC\,6384's large scale analysis. 
With large open circle we indicate a bright near nuclear source, which we investigate 
later whether it could be a star cluster or a foreground star. 

\begin{figure}
\includegraphics[width=0.48\textwidth]{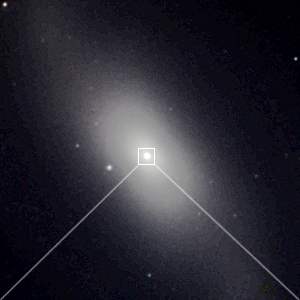}
\includegraphics[width=\columnwidth,bb=25 0 335 378]{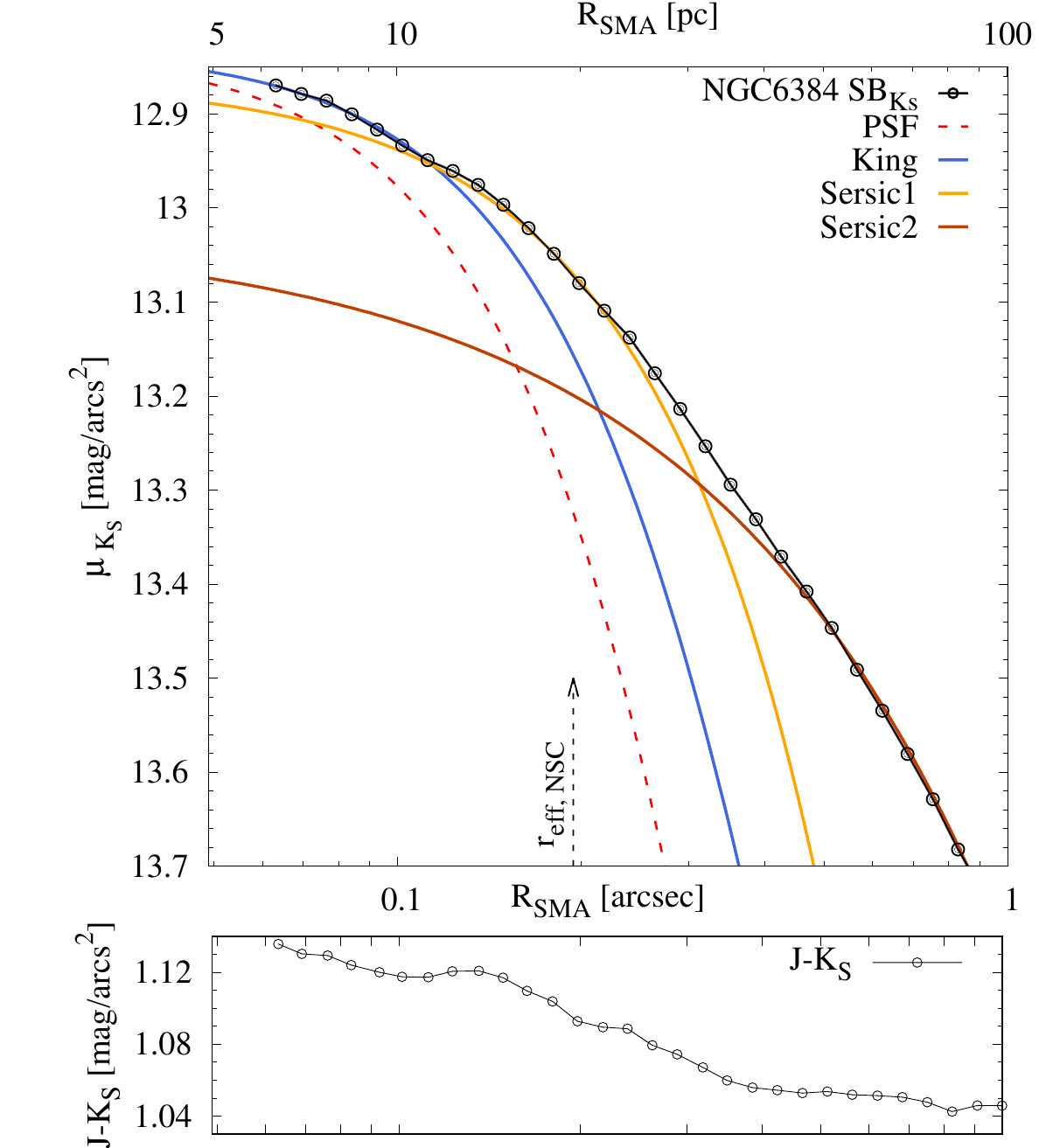}
\caption{{\bf Top: }$JHK_S$ image of NGC\,6384 central $3.6\!\times\!3.6$\,kpc 
($\sim\!36\arcsec\!\times\!36\arcsec)$. {\bf Middle: }$K_S$ surface brightness 
profile  (line with circles) of the nuclear $r\!=\!1\arcsec$ (100\,pc). 
Various components obtained from 2D fitting (see figure legend) illustrate the resolved nature 
of the NSC. {\bf Bottom: } $J\!-\!K_S$ surface colour profile.
\label{fig:SB}
}
\end{figure}

The integrated magnitude and colour of the NSC is shown with asterisk in Fig.\,\ref{fig:CMD} 
(details on the NSC photometry in \S\,\ref{Sect:Struct:NSCGal}). Before we model 
the NSC SED in \S\,\ref{Sect:SED}, here we perform a qualitative assessment of the expected NSC 
mass based on its luminosity. The NSC has a $J-$band luminosity of $\sim4\times10^7L_{J,\odot}$ 
(cf. Fig.\,\ref{fig:CMD}), which will correspond to a mass of $\sim\!2\times10^7M_\odot$, 
for a $M/L_J\!\sim\!0.4$ increasing by $\sim\!0.1$ from low to high metallicity estimated 
from \cite{BC03} SSP models for an old ($t\simeq5$\,Gyr) stellar population. For a 14\,Gyr 
old SSP, the M/L is roughly 0.9, i.e. it would be a factor of two more massive. For 
comparison, the Galactic NSC has a mass of the same order \cite[$2.4\times10^7M_\odot$][]{Schoedel07,Schoedel14,Feldmeier14}. 
We also note that the nuclear star cluster of NGC\,6384 shows a fairly red $J\!-\!K_S$ 
colour compared to the rest of the sources and the SSP tracks in the CMD. Such a colour 
is, however, typical for a galactic nuclear source, which can be composed by a mixture 
of old and young stellar populations and more importantly, could be highly reddened. 
Therefore, to lift various degeneracies in \S\,\ref{Sect:SED} we combine our NIR with 
the HST optical data to perform a detailed SED analysis of the NSC stellar population 
composition and reddening. 

\subsection{Structure and photometry of the NSC and NGC\,6384}\label{Sect:Struct:NSCGal}

\begin{figure}
\includegraphics[width=0.495\columnwidth]{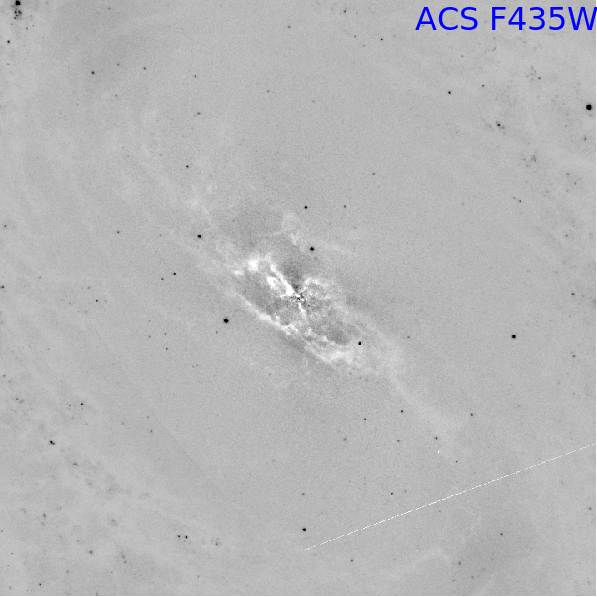}
\includegraphics[width=0.495\columnwidth]{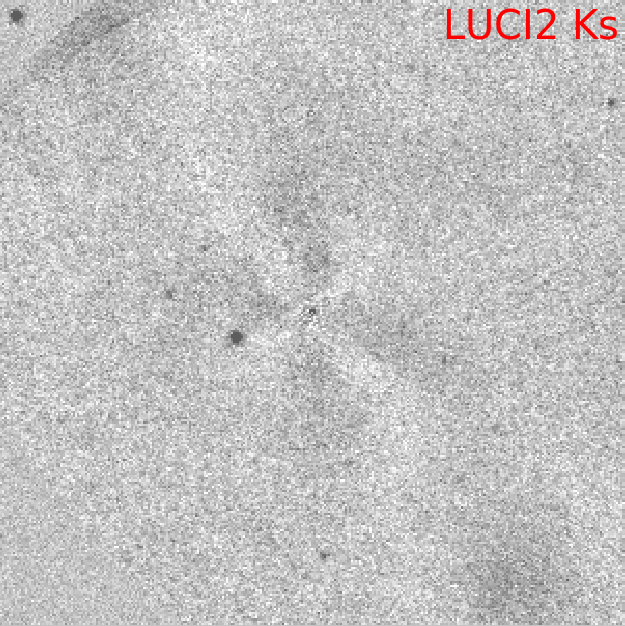}
\caption{Residual images from the subtraction of an elliptical model. The dusty disk/lane 
seen in the optical/blue $F435W$ band (left) is, expectedly, completely transparent in the 
NIR $K_S$ filter. Pixel masks for all filters were con\-structed from these residual images 
and are used for multi-component~2D mo\-delling. Both panels have identical size of 
$3.6\times3.6$\,kpc ($36\arcsec\!\times\!36\arcsec)$.
\label{fig:ellipse_rsd}}
\end{figure}

\begin{figure}
\includegraphics[width=0.48\textwidth]{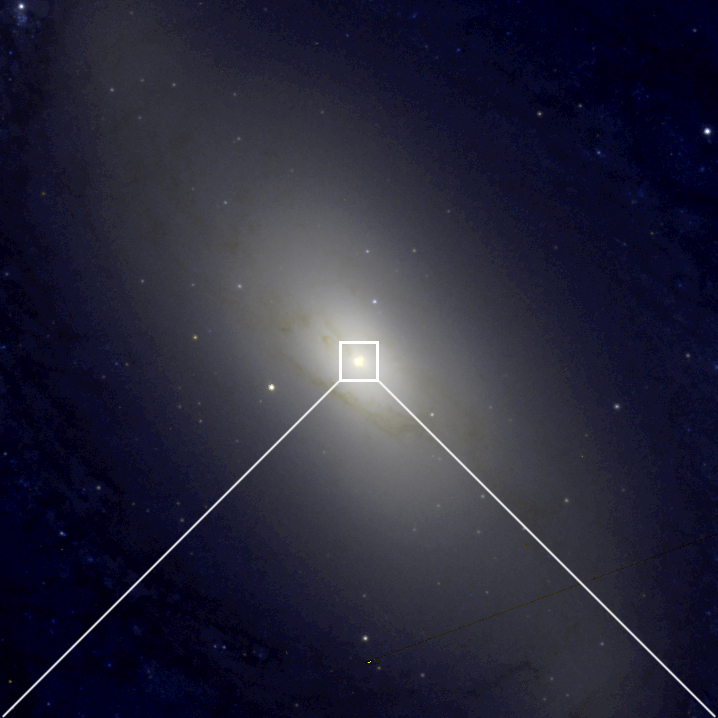}
\includegraphics[width=0.48\textwidth,bb=30 0 342 378]{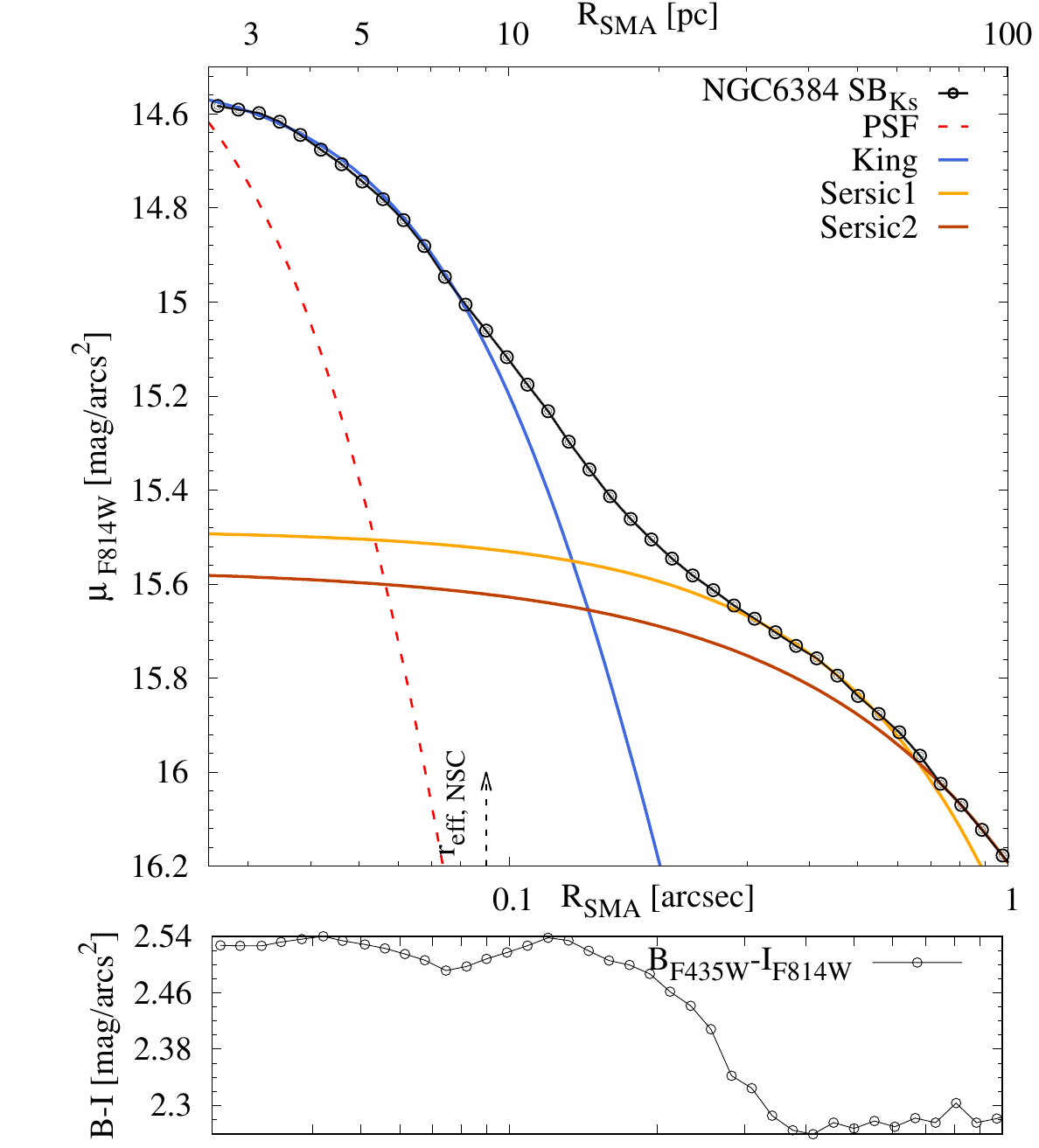}
\caption{{\bf Top:} Two filter ($F435W,\ F814W$) HST/ACS colour composite image of NGC\,6384 of the 
same area as in Fig.\,\ref{fig:SB} and \ref{fig:imfit_rsd}. Contrast and scaling are adjusted to 
clearly see the NSC and the dusty disk-like feature. {\bf Middle:} $I$-band $(F814W)$ SB and {\bf bottom:} 
$B\!-\!I$ colour profiles of the indicated area. Magnitudes are corrected for foreground Galactic 
reddening only.
}\label{fig:ACScolour}
\end{figure}

\begin{figure}
\includegraphics[width=\linewidth]{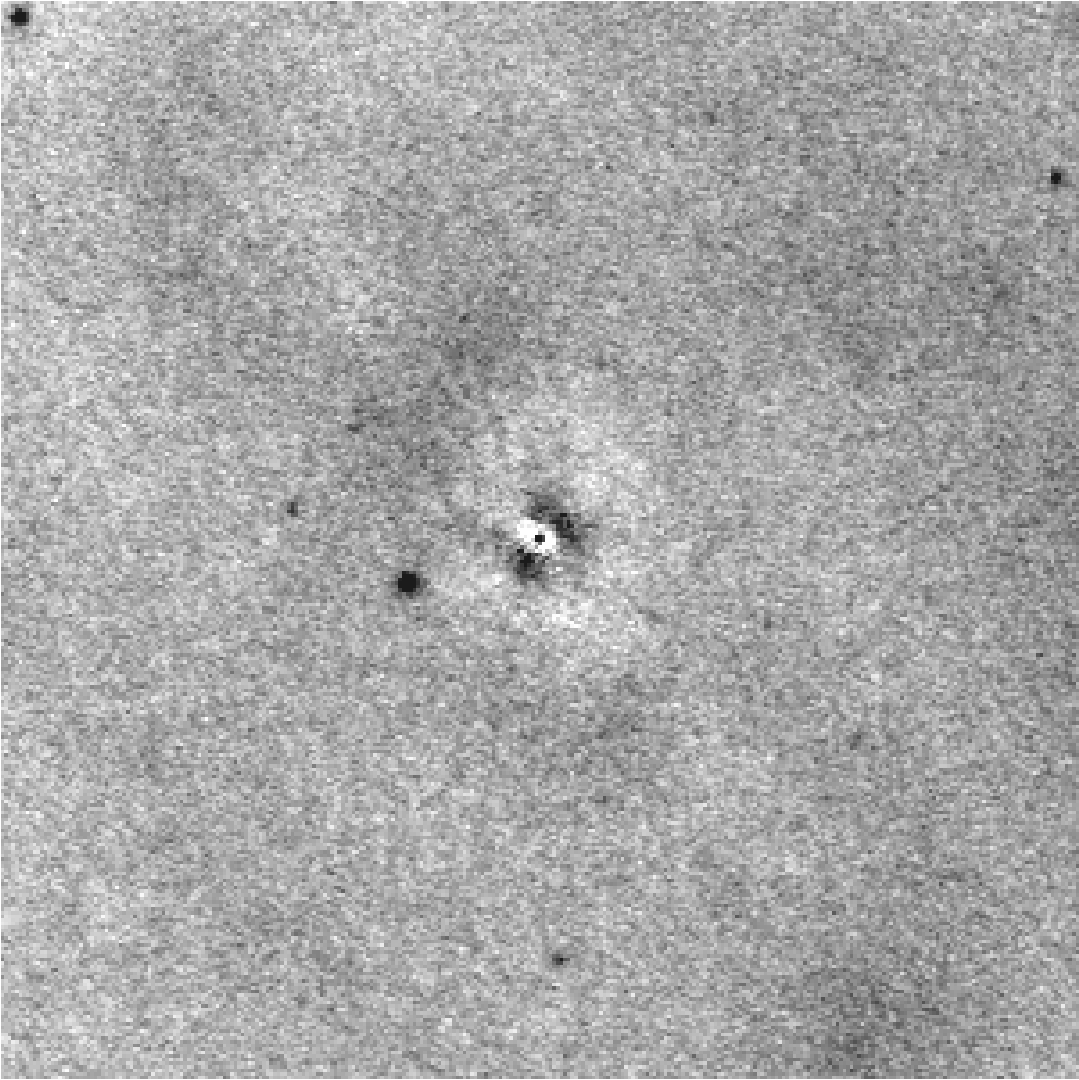}
\caption{$K_S$-band residual image from a model fit either without the inner S\'ersic ($r_{\rm eff}\!\sim100$\,pc$\simeq\!1\arcsec$) 
or the outer S\'ersic ($r_{\rm eff}\!\sim\!400$\,pc$\simeq\!5\arcsec$) com\-po\-nents. The result is 
identical for the other filters and the HST images as well.
}\label{fig:without_innersersic}
\end{figure}

As it can be seen in Figure\,\ref{fig:N6384_JHKs_colour}, NGC\,6384 has a boxy bulge. 
An indication for an X-shape is better seen in the residual $K_S-$band image in 
Figure\,\ref{fig:ellipse_rsd}\,b, which is similar to that of the MW \cite[e.g.][]{Dwek95,McWilliam&Zoccali10,Ness&Lang16,Abbott17} 
and M\,31 \citep{Beaton07,Athanassoula06}, as well as in many other galaxies \cite[e.g.][]{Laurikainen11,Erwin&Debattista17}. 
The boxy bulge was also quantified from seeing limited WHT/INGRID $K_S$-band data by 
\cite{Erwin&Debattista13}, who expectedly were unable to resolve the nuclear cluster. 
\cite{Georgiev&Boeker14} analysed the HST/WFPC\,2 $F606W$ image of the NSC and fitted 
its profile with a single King \citep{King62} model using {\sc ishape} and obtained 
a half-light radius of $r_{\rm eff}\!=\!15.9_{-0.0}^{+0.5}$\,pc. Here we re-analyse that 
data too to account for the other structures that might bias the $r_{\rm eff}$, especially 
in the inner $1\arcsec-2\arcsec$. In particular, we focus on the central $3.6\!\times\!3.6$\,kpc ($\sim\!36\arcsec\!\times\!36\arcsec)$ 
region of NGC\,6384, which covers the bar, the boxy bulge and parts of the galaxy disk 
(Fig.\,\ref{fig:SB} top). We focus only on this area, because it is sufficient for the analysis of the NSC and it 
is fully covered by the archival HST/WFPC\,2 high-resolution (0.05\arcsec/pix) PC\,1 chip, 
and the ACS camera drizzled to identical resolution (see \S\,\ref{Sect:DataReduction.HST}). 
For all images, we first perform a simple isophotal fitting with the {\sc ellipse} task 
in IRAF/PyRAF, which gives a good idea for various substructures (e.g. dust lanes) when 
we subtract the fitted isophotal model from the image. We note that the {\sc ellipse} 
task does not account for the PSF during the SB extraction, therefore, the magnitude and 
colour SB-profiles shown here are merely for illustration purposes and should not be 
considered for quantitative analysis by the reader for radii smaller or comparable to the 
PSF radius ($r\!\lesssim\!0.12\arcsec$). We show an example of the residual images from 
subtracting the {\sc ellipse} model in Figure\,\ref{fig:ellipse_rsd}, for the two filters more and 
less affected by extinction: the ACS/$F435W$ (B) and the LUCI\,2 $K_S$, respectively. 
It is evident that there is a prominent dusty disk/lane swirling around the nuclear zone 
($2.5\arcsec\times5\arcsec\simeq250\times500{\rm pc}$), which expectedly is completely 
transparent in the NIR $K_S$-band. From these images for all filters we create a pixel 
mask, which we use in a next iteration with {\sc ellipse} and later on during the 2\,D 
fitting. In Figure\,\ref{fig:SB} (middle) we present the $K_S$-band 1\,D surafce brightness 
(SB) profile (line connected open circles), because it is least 
affected by extinction, has the sharpest PSF of our NIR data and traces the old stellar 
population that dominates the mass. In the figure we only show the inner $r\!=\!1\arcsec\ (\sim100$\,pc) 
to illustrate the high spatial resolution achieved with ARGOS and the resolved nature 
of the NSC. Clearly, the NSC is well resolved beyond the PSF (dashed, red line in 
Fig.\,\ref{fig:SB}) and it is well represented by a King profile (solid, blue curve), 
whose parameters we derived as explained below. A second component immediately surrounding 
the NSC is also clearly seen at $r\!\sim\!0.2\arcsec\simeq\!15$\,pc, which is featured 
by a small bump in $J\!-\!K_S$ colour shown in the bottom panel of Figure\,\ref{fig:SB}. 
This feature is less prominent in the SB-profile from the HST/ACS $F814W$ image shown 
in Figure\,\ref{fig:ACScolour}, which is likely pointing to the fact that it is an old 
structure or/and more obscured in the optical. However, we see that a different component 
is becoming more prominent starting at around $\sim0.3\arcsec$, which suggests that it 
must be composed of younger stellar population. To avoid over interpretation of the 
colour profiles shown in Figures\,\ref{fig:SB} and \ref{fig:ACScolour} we note at this 
point that due to the fact that {\sc ellipse} does not account for the PSF, the apparent 
marginal trend of redder colour (if any) in the core $r\!\lesssim\!0.12\arcsec$ are mainly 
due to the sharper PSF in the redder filter. We further discuss these components and 
colour trends in \S\,\ref{Sect:Discussion}. The point is that in order to extract correctly 
the properties of the NSC, we need to simultaneously fit for all these components in 2\,D 
and on larger scale in all filters with their respective PSFs.

\begin{figure*}
\includegraphics[width=0.33\textwidth]{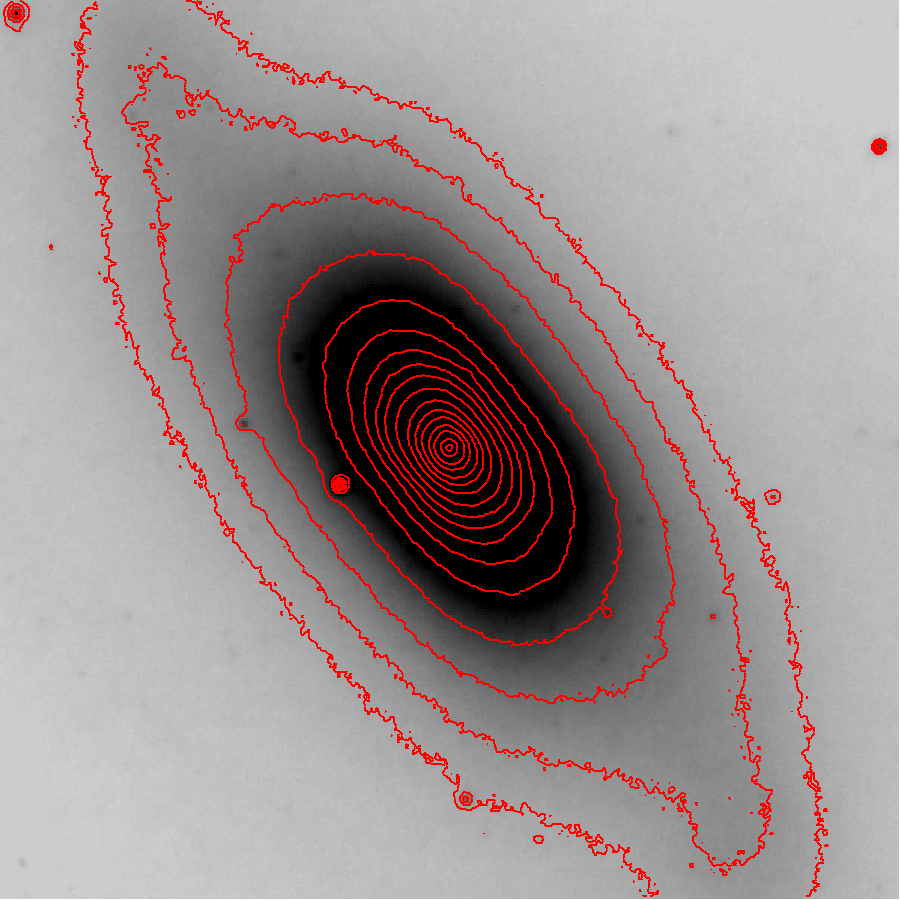} 
\includegraphics[width=0.33\textwidth]{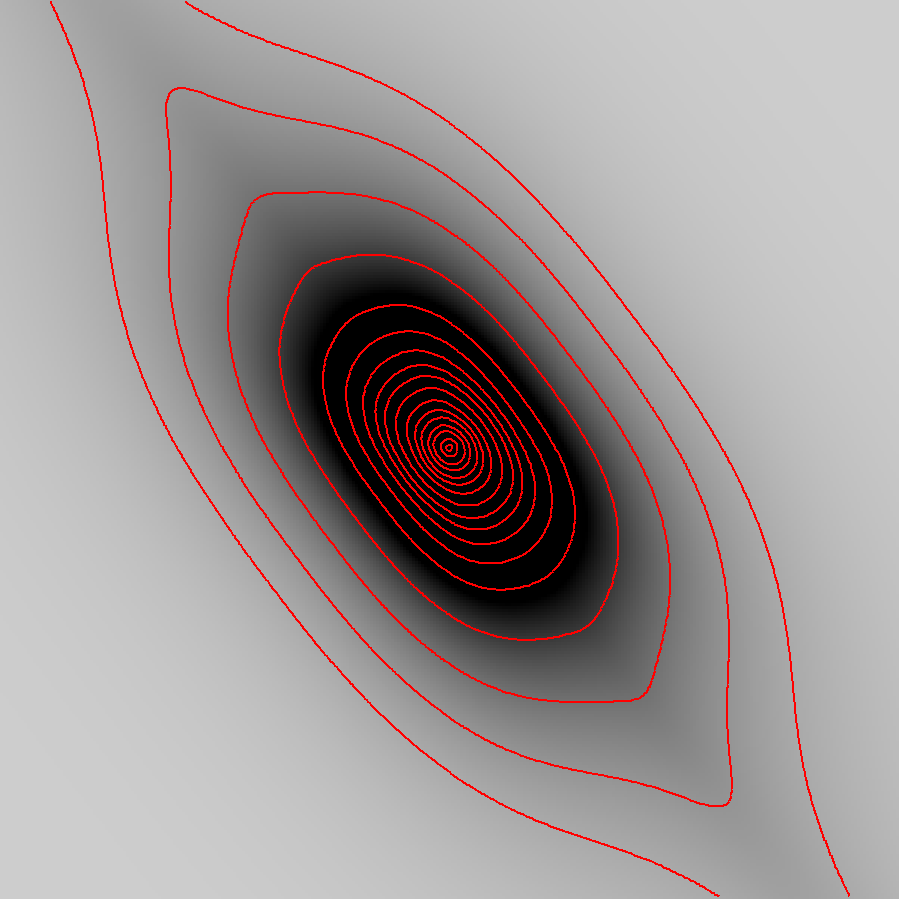} 
\includegraphics[width=0.33\textwidth]{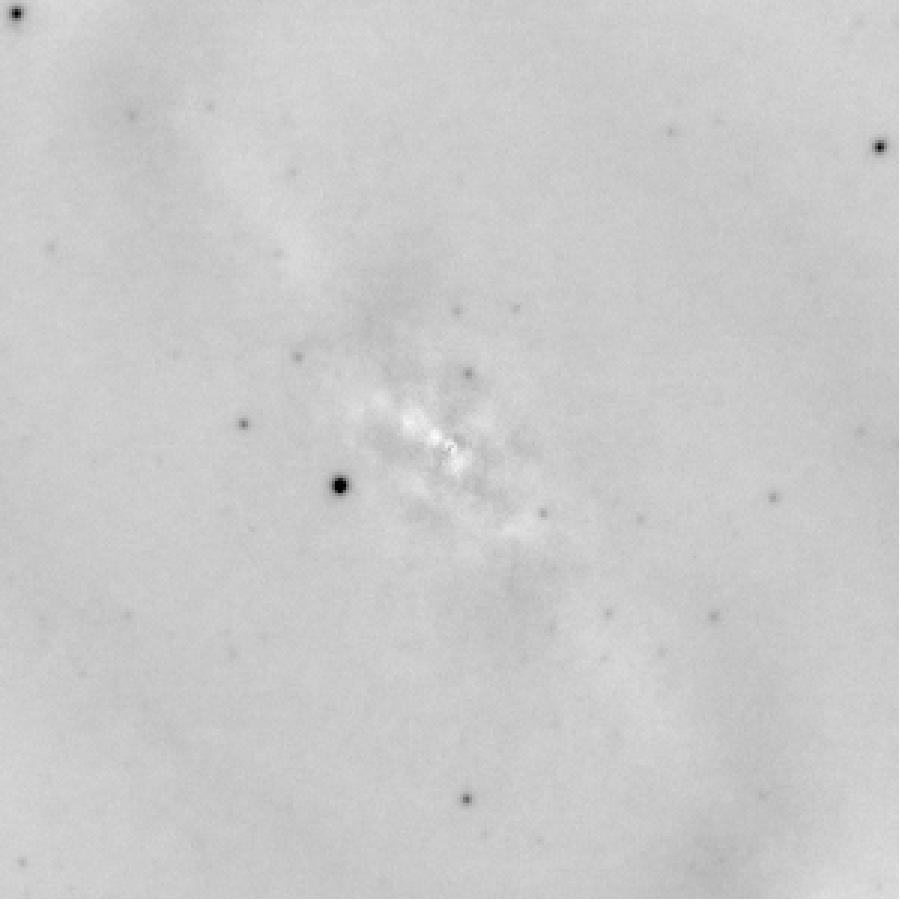} 
\includegraphics[width=0.33\textwidth]{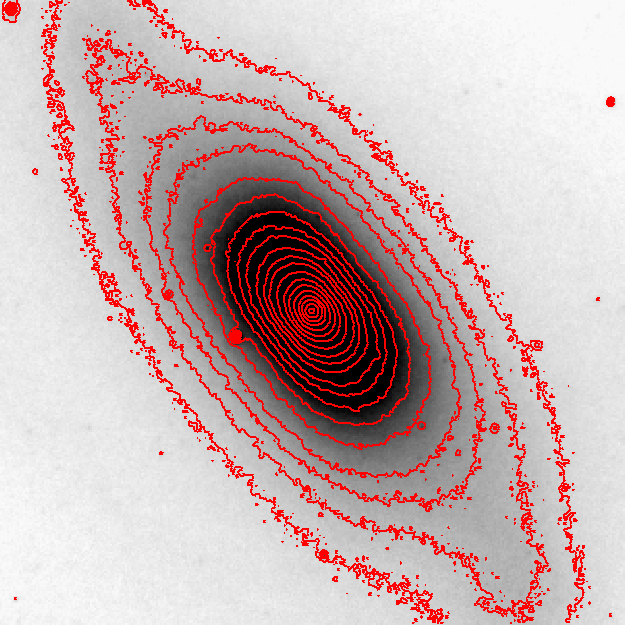} 
\includegraphics[width=0.33\textwidth]{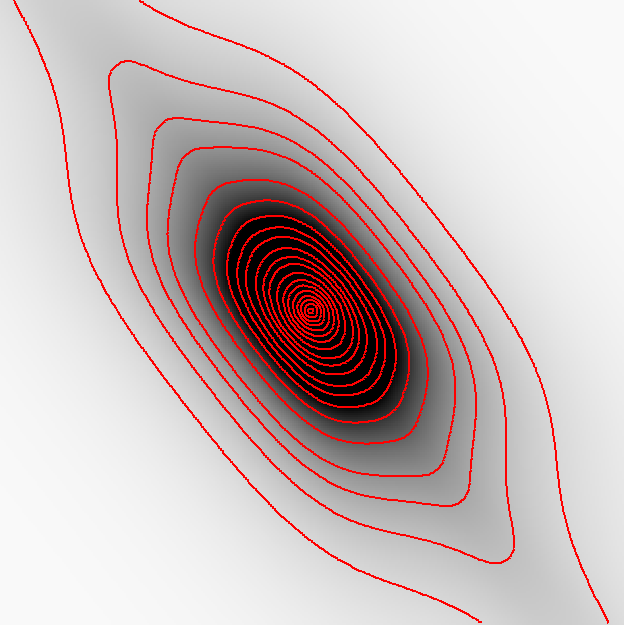} 
\includegraphics[width=0.33\textwidth]{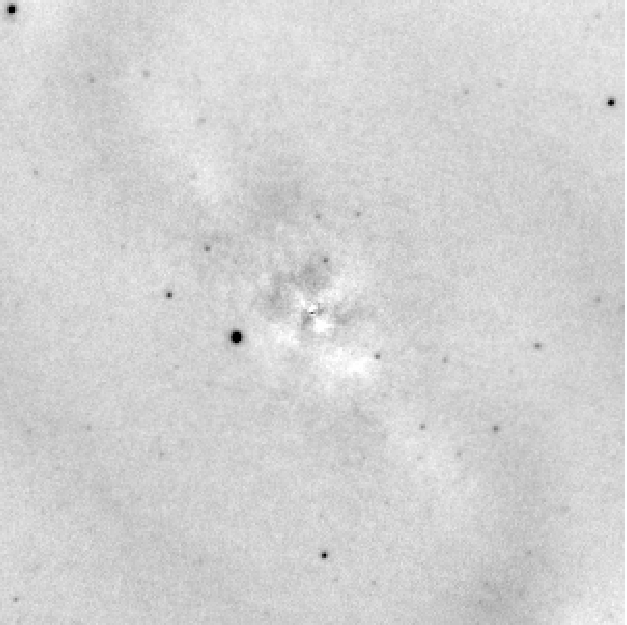} 
\includegraphics[width=0.33\textwidth]{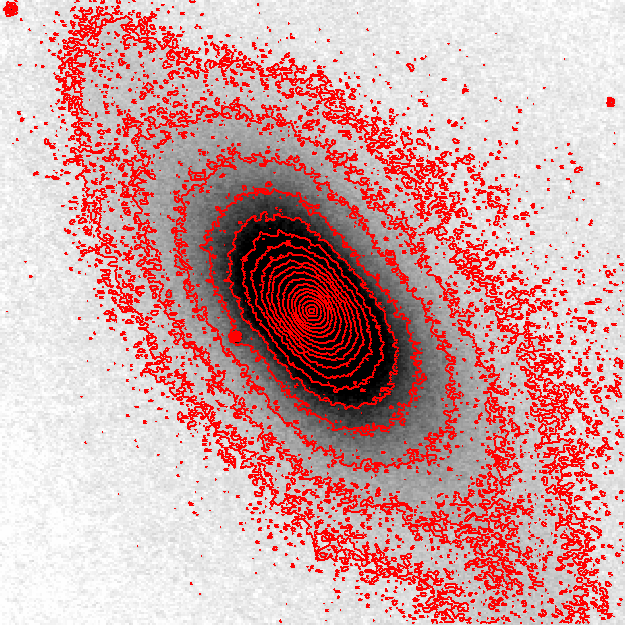}
\includegraphics[width=0.33\textwidth]{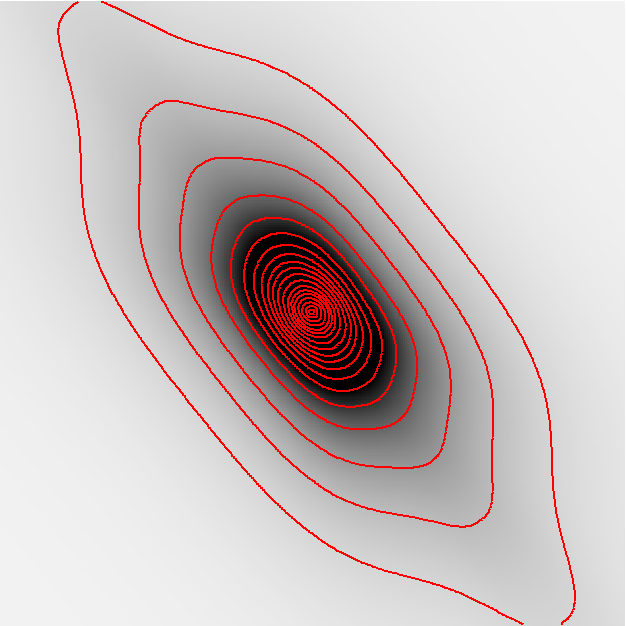}
\includegraphics[width=0.33\textwidth]{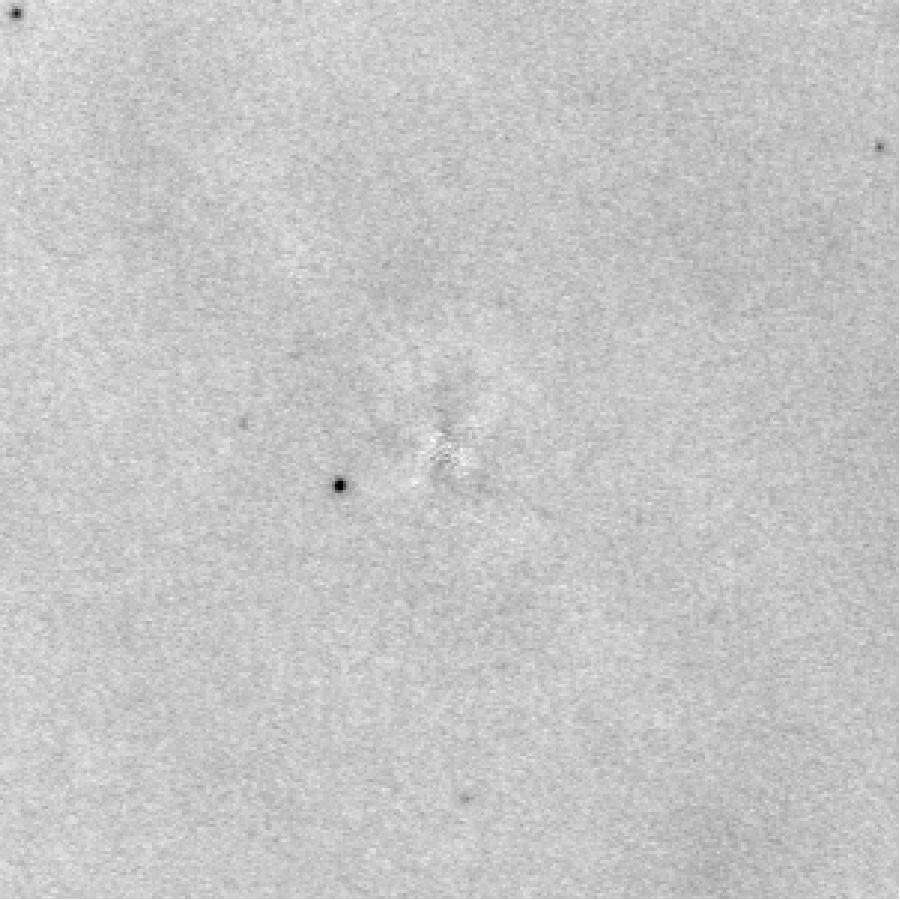}
\caption{{\bf Top to bottom:} $J,\ H,\ K_S$ NGC\,6384 images ({\bf left column}) of the central 
$3.6\!\times\!3.6$\,kpc ($\sim\!36\arcsec\!\times\!36\arcsec)$. {\bf Middle:} Best fitting model 
and {\bf right column} is the residual image (model minus the data). For reference and to guide 
the eye, we show the with red contours the data and model images at same isophotal levels.
\label{fig:imfit_rsd}
}
\end{figure*}

\begin{table*}
\begin{minipage}{\textwidth}
\footnotesize
\caption{Best fit parameters for the NSC and the inner 1\,kpc of NGC\,6384. Each table block gives the parameter values of the respective fitted profile type.
\label{Table:bestfits}
}
%\begin{tabular}{p{0.55cm}p{1cm}p{1cm}p{1cm}p{1cm}p{1cm}p{1cm}}
\begin{tabular}{lllllll}
\hline\hline
\hspace*{-.2cm}Profile &
$B_{F435W}$ &
$V_{F606W}$ &
$I_{F814W}$ &
$J$ &
$H$ &
$K_S$ \\
\hspace*{-.2cm}(1) &
(2) &
(3) &
(4) &
(5) &
(6) &
(7) \\
\hline
\multicolumn{1}{l}{\bf King (the nuclear star cluster)} \\
%&&&\\
\hspace*{-.2cm}Mag\footnote{Magnitudes are corrected for Galactic reddening only. Intrinsic self-absorption values are obtained via SED fitting in \S\,\ref{Sect:SED}.} & 21.61 & 19.98&18.56&16.39&15.25&14.70\\
\hspace*{-.2cm}$r_{\rm eff}$\,[pc]\footnote{Size given in pc or kpc is for an assumed distance of 20.7\,Mpc. We calculate $r_{\rm eff},\ {\rm from\ }\ r_c,\ c,\ \alpha$ of the modified King model as given in \S\,\ref{Sect:App.r_eff_King}. } &$5.8$&$10.1$&$9.1$&$13.9$& $11.8$ & $19.5$\\
\hspace*{-.2cm}$r_c$\,[pc]	& $0.71^{+0.5}_{-0.3}$ & $1.59^{+0.4}_{-0.4}$& $1.94^{+0.01}_{-0.4}$&$2.87^{+0.10}_{-0.10}$ & $2.32^{+1.1}_{-0.4}$ &$4.26^{+0.09}_{-1.0}$ \\
\hspace*{-.2cm}$C$	& $146.6^{+3.1}_{-0.2}$ & $155.02^{+0.7}_{-21.7}$ &$86.4^{+8.5}_{1.6}$&$97.27^{+3.6}_{-2.9}$ &$87.5^{+13.04}_{3.1}$ &$74.0^{+7.1}_{-6.8}$ \\
\hspace*{-.2cm}$\alpha$	&	$1.28^{+0.28}_{-0.28}$ & $1.98^{+0.40}_{-0.40}$&$2.08^{+0.58}_{-0.18}$&$2.13^{+0.08}_{-0.10}$&$1.86^{+0.21}_{-0.10}$&$1.94^{+0.98}_{-1.08}$\\
\hline
\multicolumn{1}{l}{\bf Sersi\'c (disk/bulge?)}\\
\hspace*{-.2cm}Mag & 17.30 & 16.00&15.35&14.21&13.59&13.22\\
%\hspace*{-.2cm}PA       &      $65.0^{+0.02}_{-0.02}$ & $0.0^{+0.02}_{-0.02}$&$37.17^{+0.07}_{-0.03}$& $0$&$0$& $0$\\
%\hspace*{-.2cm}ell       &      $0.24 & $0.24&$0.15^{+0.09}_{-0.09}$& $0$&$0$& $0$\\
\hspace*{-.2cm}n        & $0.78^{+0.01}_{-0.01}$& $0.76^{+0.01}_{-0.01}$&$0.79$&$0.724$&$0.716$&$0.65^{+0.01}_{-0.01}$\\
\hspace*{-.2cm}$r_{\rm eff}$\,[pc] & $102.1^{+0.2}_{-0.2}$ & $106.2^{+0.5}_{-0.33}$&$100.3^{+0.1}_{-0.3}$& $106.44^{+0.2}_{-0.2}$& $101.15^{+0.11}_{-0.11}$& $102.26^{+5.11}_{-5.0}$\\
\hline
\multicolumn{1}{l}{\bf Sersi\'c (disk/bulge?)}\\
\hspace*{-.2cm}Mag & 15.04 & 13.59&13.55&13.01&11.73&11.47\\
\hspace*{-.2cm}PA\footnote{If PA or ellipticity is not given, their values are 0. If a parameter has no uncertainty, its value is below 1\%.} & $31.53^{+0.04}_{-0.04}$& $31.0^{+0.03}_{-0.03}$&$30.22^{+0.03}_{-0.03}$& $33.9^{+0.03}_{-0.03}$&$36.46^{+0.02}_{-0.02}$& $32.59^{+0.11}_{-0.11}$\\
\hspace*{-.2cm}ell       &      $0.42$ & $0.51$&$0.45^{+0.01}_{-0.02}$& $0.48$&$0.46$& $0.45$\\
\hspace*{-.2cm}n         &      $0.97$ & $1.13$&$0.99$& $1.044$&$1.214$& $1.102$\\
\hspace*{-.2cm}$r_{\rm eff}$\,[pc] & $410.4^{+0.5}_{-1.3}$ & $669.2^{+2.5}_{-1.1}$&$490.9^{+0.5}_{-0.6}$& $465.4^{+1.0}_{-1.0}$&$421.5^{+0.3}_{-0.3}$& $396.2^{+3.6}_{-5.6}$\\
\hline
\multicolumn{1}{l}{\bf Sersi\'c GenEllipse (boxy bulge)}\\
\hspace*{-.2cm}Mag & 14.85 & 13.06&11.39&10.32&9.84&9.35\\
\hspace*{-.2cm}PA       &      $37.47^{+0.02}_{-0.02}$ & $37.05^{+0.02}_{-0.02}$&$37.18^{+0.07}_{-0.03}$& $37.79^{+0.17}_{-0.3}$&$38.40^{+0.3}_{-0.8}$& $38.98^{+0.05}_{-0.05}$\\
\hspace*{-.2cm}ell       &      $0.35$ & $0.351$&$0.39^{+0.09}_{-0.09}$& $0.40$&$0.39^{+0.11}_{-0.04}$& $0.40$\\
\hspace*{-.2cm}c0         &      $1.90^{+0.01}_{-0.01}$ & $1.217^{+0.01}_{-0.01}$&$0.68^{+0.01}_{-0.0}$& $0.784^{+0.38}_{-0.29}$&$1.40^{+0.56}_{-0.57}$& $1.117^{+0.02}_{-0.02}$\\
\hspace*{-.2cm}n         &      $0.52$ & $1.35^{+0.01}_{-0.01}$&$1.83^{+0.01}_{-0.0}$& $1.613^{+0.36}_{-0.3}$&$1.36^{+0.23}_{-0.14}$& $1.49^{+0.01}_{-0.01}$\\
\hspace*{-.2cm}$r_{\rm eff}$\,[pc] & $788.6^{+0.5}_{-0.5}$ & $776.8^{+2.66}_{-2.38}$&$1182.4^{+0.6}_{-0.5}$& $925.99^{+10.7}_{-4.5}$&$937.3^{+0.1}_{-0.1}$& $1021^{+4.1}_{-3.9}$\\
\hline
\multicolumn{1}{l}{\bf Exponential GenEllipse (disk)}\\
\hspace*{-.2cm}Mag & 13.66 & 11.84&11.71&10.17&9.62&10.35\\
\hspace*{-.2cm}PA       &      $35.2$ & $38.464$ & $36.39$& $37.95^{+0.1}_{-0.1}$&$38.4^{+0.06}_{-0.04}$& $38.65^{+0.04}_{-0.04}$\\
\hspace*{-.2cm}ell       &      $0.75$ & $0.813$&$0.96$& $0.94^{+0.08}_{-0.36}$&$0.94^{+0.10}_{-0.17}$& $0.902$\\
\hspace*{-.2cm}c0         &      $-1.05^{+0.01}_{-0.01}$ & $-1.028$&$2.07^{+0.04}_{-0.03}$& $1.27^{+0.02}_{-0.08}$&$1.07^{+0.62}_{-0.08}$& $1.278^{+0.161}_{-0.139}$\\
\hspace*{-.2cm}$h$\,[kpc]\footnote{We caution that due to fitting only the inner 1\,kpc, the exponential disk scale height might not be truly representative for the actual extent of the disk.} 
& $2.86^{+0.01}_{-0.01}$& $4.68^{+0.01}_{-0.01}$&$5.92^{+0.01}_{-0.01}$& $6.58^{+0.05}_{-0.15}$&$6.38$&$2.65^{+0.73}_{-0.70}$\\

\hline\hline
\end{tabular}
%\footnotetext{a}
\end{minipage}
\end{table*}

We model the light profile of the central $3.6\!\times\!3.6$\,kpc ($\sim\!36\arcsec\!\times\!36\arcsec)$ 
of NGC\,6384 and its NSC with {\sc imfit}\footnote{http://www.mpe.mpg.de/$\sim$erwin/code/imfit/} 
\citep{Erwin15}. This software package performs a 2D profile fitting using the image PSF, 
a large choice of analytical profiles, iterative fitting minimization techniques ($\chi^2$, 
maximum-likelihood), bootstrapping and MCMC modules for exploring and deriving more reliable 
values of the fitted parameters and their uncertainties. The high $S/N$ of the NSC, allows 
us to use a $10\times$ oversampled PSF to fit the central 10\,x\,10 pixels 
($\sim\!1.2\arcsec\simeq\!120$\,pc). We experimented with a wide range 
of analytical models available in {\sc imfit} and found that a five component model described 
best the fitted area (see Table\,\ref{Table:bestfits}). Namely, the NSC at the very centre 
was best fit by a modified \cite{King62} model with concentration, core radius and $\alpha$ as 
fitted parameters \citep{Elson99,Ch.Peng10}. The inner $r\!\lesssim\!2\arcsec\simeq\!200$\,pc 
required two S\'ersic components for the fit around the NSC (cf. Fig.\,\ref{fig:SB}), while the 
large scale boxy bulge and disk were fitted with generalized elliptical S\'ersic and Exponential 
profiles \cite[see ][and {\sc imfit} manual]{Erwin15}. We experimented with other functions, 
including nuclear ring and disk, pure Gaussian and Core-S\'ersic models, however, we achieved 
worse fits. It was clear from this exploration that the aforementioned components gave the 
smallest residuals and best fit values compared to choosing others or using less or more fitting 
functions. To illustrate the need for using more components, in Figure\,\ref{fig:without_innersersic} 
we show the $K_S$-band residual image fitted either without the inner S\'ersic ($r_{\rm eff}\sim100$\,pc$\simeq1\arcsec$) 
or the outer S\'ersic ($r_{\rm eff}\sim400$\,pc$\simeq5\arcsec$). It is evident that strong 
residuals arise due to averaged profile resulting from the fit, which highlights the need 
for having both. Finally, we used the {\sc imfit-mcmc} module to explore a wider region for 
the range of the fitted parameters via MCMC. From the posterior distributions we obtained their 
best values and uncertainties. An example is shown in Appendix\,\ref{Sect:App.imfit_mcmc}, 
Figure\,\ref{fig:imfit_mcmc} for the $K_S$-filter. The results from fitting NGC\,6384 inner 
3.6\,kpc in the $JHK_S$ filters are illustrated in Figure\,\ref{fig:imfit_rsd} and the best 
fit parameter values for all, including ACS and WFPC\,2 filters, are given in Table\,\ref{Table:bestfits}. 
The different blocks in the table indicate the different components. The left column of 
Figure\,\ref{fig:imfit_rsd} shows the image in each band (from top to bottom), in the middle 
is the best fit model and in the right is the residual image (model minus data). 
The dust lane is completely transparent in the $K_S$ image (Fig.\,\ref{fig:imfit_rsd}\,bottom-right), 
becomes more opaque with decreasing wavelength toward $H$ and $J$-bands and is well visible in 
the optical ACS filters (cf. Fig.\,\ref{fig:ellipse_rsd}, left). This dusty, disk-like structure 
is not drastically obscuring the NSC, as it can be seen in Figure\,\ref{fig:ACScolour}\,(top), 
to require severe pixel masking that can hamper the fit of the optical profile 
(cf Fig.\,\ref{fig:ACScolour}\,middle). This is further supported by the lack of strong colour 
gradient (cf Fig.\,\ref{fig:ACScolour}\,bottom). The colour and its gradient as well as the 
structure of the various components given in Table\,\ref{Table:bestfits} is discussed in 
\S\,\ref{Sect:Discussion}.

\subsection{SED analysis of the central kiloparsec}\label{Sect:SED}

Here we analyse the SED of NGC\,6384 of the overlapping region between our NIR images 
and the archival HST optical data. To closely account for the extended nature and 
evidently varying structure of the NSC as a function of wavelength, we use its six filter 
model magnitudes, which unlike fixed aperture magnitudes, should not only account for 
the varying PSF with wavelength, the varying structure of the cluster, but also minimize 
contamination from flux coming from the other underlying galactic components. We also 
model the SED of the SB profile of the inner kiloparsec of NGC\,6384 in the six filters, 
which should well represent its stellar population properties on scales 
$r\!\gtrsim\!0.12\arcsec$ as PSF effects are not taken into account by the Ellipse extraction 
(see \S\,\ref{Sect:Struct:NSCGal}).

\begin{figure}
\includegraphics[width=1.02\linewidth,bb=10 0 404 300]{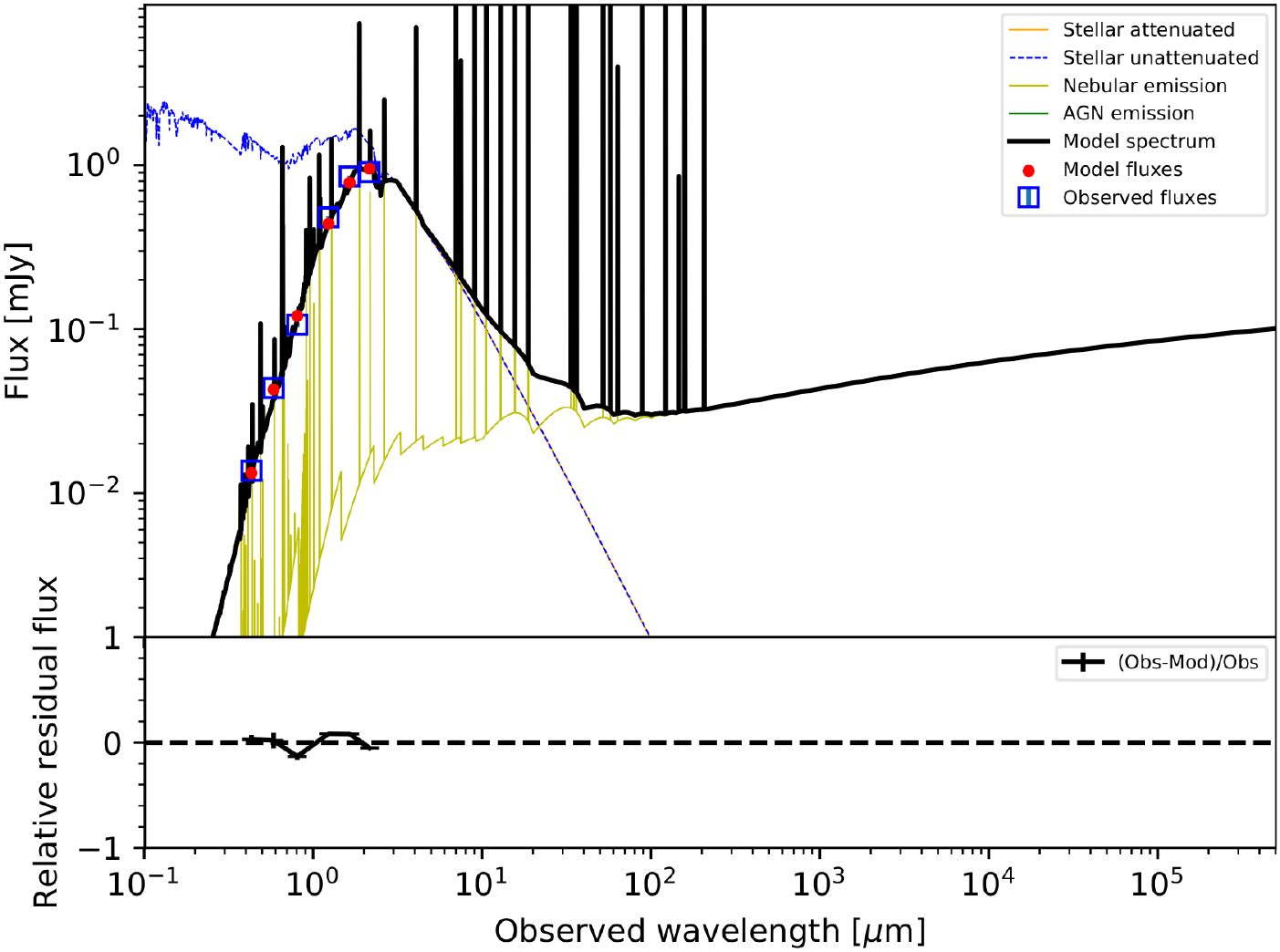}
\caption{The best SED fit to the NGC\,6384 nuclear star cluster using six-band optical-NIR 
photometry (HST: $F450W, F606W, F814W$ and LBT/LUCI: $J,H,K_S$). The fluxes are from the 
model magnitudes in Table\,\ref{Table:bestfits} obtained via 2\,D image modelling (in 
\S\,\ref{Sect:Struct:NSCGal}). SED fitting details in \S\,\ref{Sect:SED}.
}\label{fig:NSC_SED}
\end{figure}

\begin{table}
\caption{Properties of the nuclear star cluster from SED fitting (in \S\,\ref{Sect:SED}) to 
the six filter optical-NIR model magnitudes in Table\,\ref{Table:bestfits}.
\label{Table:SED}
}
\begin{tabular}{p{.5cm}p{1.55cm}p{1.55cm}p{1.55cm}c}
\hline\hline
Population & \centering Age & \centering Metallicity, [M/H] & \centering $E(B-V)$ & Mass\\
&\centering [Gyr] & \centering [dex] & \centering [mag] & [$10^6M_\odot$]\\
\hline
Old & \centering $10.85\pm1.33$ & $-0.11\pm0.16$ & $0.63\pm0.15$ & $35.3\pm21.6$ \\
Young & \centering $0.23\pm0.14$ & $0.33\pm0.13$ & $1.44\pm0.33$ & $2.86\pm1.9$ \\
\hline\hline
\end{tabular}
\end{table} 

Our SED fitting uses a mixture of old and young stellar populations, including nebular 
emission, fitting for dust extinction and testing for a weak AGN component. The latter 
is motivated by the fact that NGC\,6384 is similar by mass to the MW and M\,31 and its 
NSC could similarly harbour a MBH of 1-10 million solar mass, which might be more active. 
Thus, a certain fraction of the flux in the core might be coming from a weak AGN component. 
NGC\,6384 is also classified as a transitional type (T2) LINER \citep{Ho97}, which further 
support the need to probe for an AGN. To test for all these components contributing to 
the SED, we use the {\sc CIGALE}\footnote{We used version 0.12.1 \href{http://cigale.lam.fr}{cigale.lam.fr}} 
code \citep{Boquien18,Noll09,Roehlly14,Burgarella05}. It has been developed to fit the 
SED of galaxies using various models of the SFH (double exponential, delayed, periodic 
and user specified), a choice of SSP model \citep{BC03,Maraston05} with a given IMF, 
adding dust attenuation \cite[][single or double power law]{Calzetti94,Calzetti00} and 
possible nebular, AGN \citep{Fritz06,Dale14} and dust emission \citep{Casey12,Dale14,DraineLi07,Draine14}. 
We first fitted with {\sc CIGALE} the SED of the NSC using the six band model magnitudes 
given in Table\,\ref{Table:bestfits}. We assumed a double exponential SFH to test for the 
presence of more than one stellar population. Although NSC may experience a prolonged 
SFH with many episodic bursts of star formation, typically, they host two main stellar 
populations which contribute, if not to all, to the majority of their current mass and light, 
i.e. an old (metal-poorer) and a younger and more metal-rich \cite[][]{Walcher06,Rossa06,Seth06,Kacharov18}. 
Under this assumption of two main bursts of star formation, we modified the {\sc bc03} 
module of {\sc CIGALE} to allow for the old and young population to be constructed from SSPs 
with different metallicity. 

\begin{figure}
\includegraphics[width=1\linewidth,bb= 20 20 374 698]{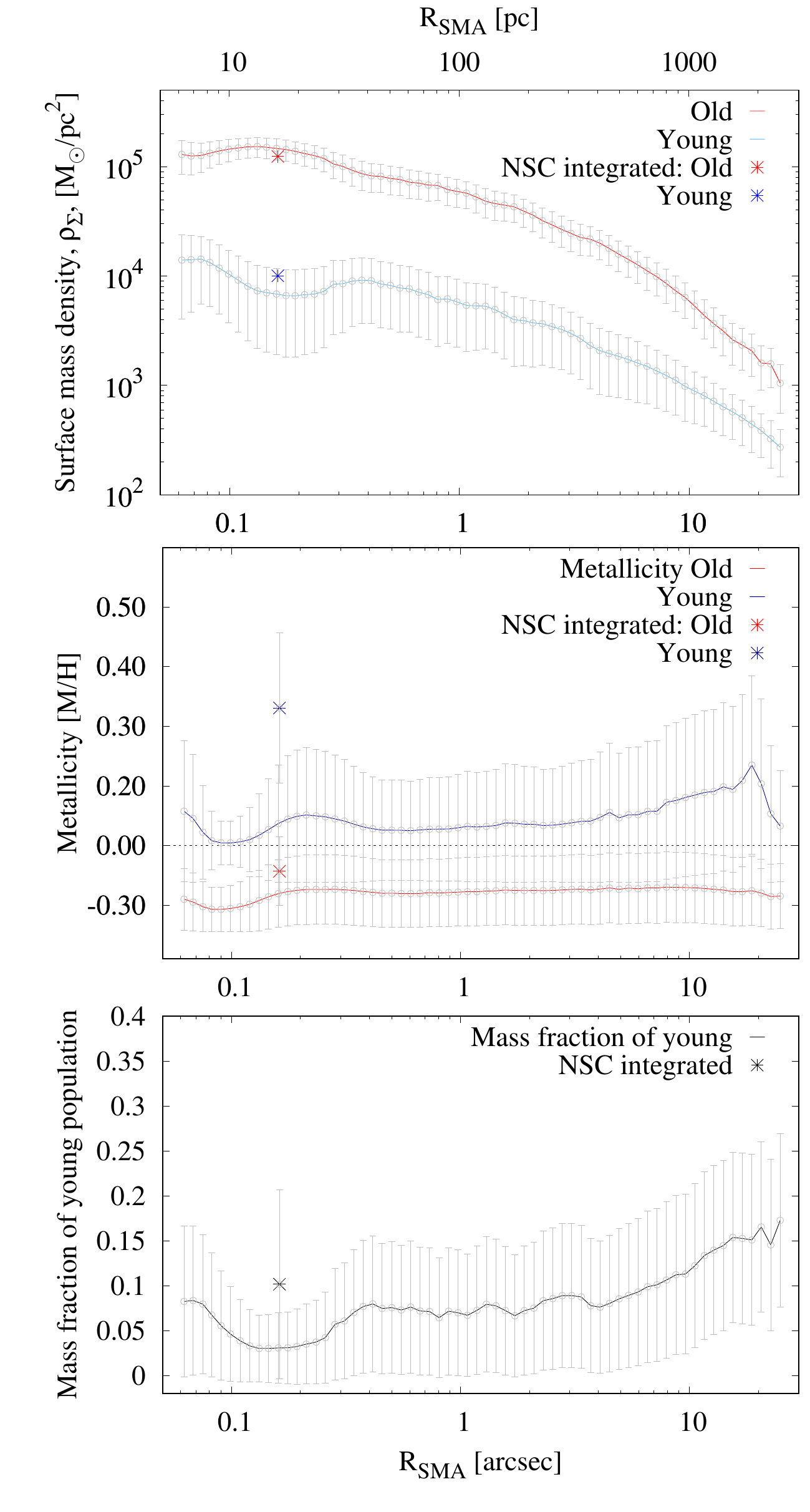} 
\caption{Radial map of the SED modelled composite stellar population. From top to bottom are 
shown the stellar mass, metallicity and mass fraction of the young stellar population.
}\label{fig:SED_map}
\end{figure}

Allowing for a wide range of possible parameter values for all aforementioned SED components, 
resulted in the synthesis of 403200 SEDs. {\sc CIGALE} analyses all these models 
compared to the observed SED in a Bayesian framework, i.e. calculating the probability 
of each model given the data, and constructing posterior distributions for each model 
parameter from which their best value and uncertainties are obtained \cite[details in ][]{Burgarella05,Boquien18}.
We show the best fit SED of the NSC in Figure\,\ref{fig:NSC_SED}, where all fitted components 
are given in the figure legend. The main properties from the six band optical-NIR SED 
fitting are summarised in Table\,\ref{Table:SED}. We did not include the AGN fraction 
of the fit into the table, as its contribution (if any) is very small, 
$<\!1\%$ ($0.46\%\pm0.36\%$), and a geometry of a minimum to maximum radii of the dusty 
torus of $r\!=\!60\pm3$ with an opening angle $\Theta\!=\!100^\circ\pm40^\circ$ and an 
angle between the AGN and the line of sight of $\psi=70.1^\circ\pm3.5^\circ$. We further 
comment on this in \S\,\ref{Sect:Discussion}.

As a consistency check, instead of using the NSC model magnitudes, we used a fixed 
aperture NSC magnitudes with a diameter of 0.35\arcsec. This is large enough to avoid 
PSF effects, contain light mostly from the NSC whose diameter is smaller than that, as 
well as it is small enough to minimize contaminating flux from the underlying structures. 
The result from fitting the NSC SED using the aperture magnitudes is that the NSC age of 
both populations remained unchanged (1\% lower); the NSC total mass decreased by $\sim\!47\%$, 
largely driven by the lower mass of the old population (by 52\%) due to the inclusion of 
contaminating flux from the likely younger underlying disk stellar population that lower 
the $M/L$. Respectively, the metallicity and attenuation values also lowered by up to 
$\sim\!45\%$. To conclude, the stellar population values for both populations obtained 
from fixed aperture are within the measurement uncertainties given in Table\,\ref{Table:SED}, 
however, as expected, there is a systematic bias due to the contaminating flux from the 
underlying disk/bulge components in direction of their respective stellar population 
properties. This result will hold true only for aperture magnitudes obtained from high 
spatial resolution observations allowing the extraction of NSC photometry from an aperture 
big enough to avoid PSF effects across different filters, but small enough to minimize 
contamination. Obviously, the latter will also strongly depend on the structure and 
profile of the underlying components.

To obtain a radial map of the stellar population properties, we also fit the SED 
at each radial location along the SB profile of all filters extracted with {\sc ellipse}. 
A summary from this fitting we show in Figure\,\ref{fig:SED_map} of the one dimensional 
projection along the semi-major axis of the main properties of the two stellar populations: 
stellar surface mass density, metallicity, and fraction of the young stellar population, from 
top to bottom, respectively. For comparison, with an asterisk in Figure\,\ref{fig:SED_map} 
we also show the integrated values of the NSC from Table\,\ref{Table:SED}, at the mean 
radial location of its effective radius from all filters of 10\,pc. All other relevant 
parameters ($E(B-V)$, age, e-folding time of the two stellar population) from the SED 
fitting did not show significant (or interesting) radial gradients. We note that although 
we extracted the radial SB beyond the 10\% of the PSF$_{\rm FWHM}$, the very central 2-3 
radial points ($\lesssim\!0.07\arcsec$) in Figure\,\ref{fig:SED_map} might not be fully 
representative and still suffer from differential PSF effects between the HST and the 
LUCI+ARGOS images.

\section{Discussion}\label{Sect:Discussion}

Here we discuss our main findings on the properties of the nuclear star cluster of NGC\,6384, 
a Milky Way like galaxy. Enabled by the sharp (0.25\arcsec) and stable PSF provided by ARGOS 
across the $4\arcmin\times4\arcmin$ LUCI\,2 field of view (\S\,\ref{Sect:App.PSF}) we analyse 
the central $3.6\!\times\!3.6$\,kpc ($\sim\!36\arcsec\!\times\!36\arcsec)$ overlap region between 
our NIR LBT and archival optical HST data. The high spatial resolution in the NIR is crucial for 
breaking age-metallicity-reddening degeneracies when combined with the optical HST data. 

\subsection{The NSC embedded in nuclear (disk) components}\label{Sect:Discussion_Multiponents}

The 2D image fitting of the high spatial resolution NIR LUCI\,2 and optical HST images allowed 
us to uncover the presence of two inner Sersi\'c components (\S\,\ref{Sect:Struct:NSCGal}, 
Fig.\,\ref{fig:SB}) with effective radii of $\sim\!100$ and 400\,pc in which the NSC with 
$r_{\rm eff}\simeq10$\,pc is embedded. We also successfully fitted for the large scale boxy 
bulge ($r_{\rm eff}\sim1$\,kpc) and disk ($\sim6$\,kpc). The properties of these multiple 
components are given in Table\,\ref{Table:bestfits}. The two inner Sersi\'c profiles have low Sersic indices 
($n\simeq0.75$ and $\simeq1$), which suggests that these might be nuclear disks. Similar 
central disks are observed in the Milky Way as well as in other galaxies. For example, stellar 
line-of-sight velocity distributions of stars in the MW reveals a nuclear disk of a truncation 
radius of $\sim$150\,pc \cite[e.g.][]{Schoenrich15}. In other galaxies, observations with 
sufficiently high spatial resolution shows that their central zones also contain nuclear disks 
\cite[e.g.][]{Balcells03,Balcells07,Seth06,Ganda09,Mendez-Abreu17}. Their formation, in the case 
of the Milky Way, is reproduced by N-body+smooth particle hydrodynamics simulations 
\citep{Debattista15,Debattista18}, which form a nuclear disk (or ring) during bar-induced 
gas inflows \citep{Cole14}. In other galaxies, similar process of star formation in nuclear 
rings is suggested to form nuclear disks \cite[e.g.][]{Kormendy&Kennicutt04}. This implies 
that the (few) hundred parsecs components around the NSC of NGC\,6384 are common in 
galaxies. However, the fact that we observe two in NGC\,6384 might hint at specific events 
in the formation past of its central regions and their eventual timing. For example, star cluster 
merger simulations do form nuclear disks of similar scales, which are also suggested to be 
useful for timing past merger events \cite[e.g.][]{Portaluri13,Sarzi16,Arca-Capuzzo16,Arca-Capuzzo17}.

\subsection{NSC effective radius varying with wavelength}\label{Sect:Discussion_reff_variat}

\begin{figure}
\includegraphics[width=1\linewidth,bb= 12 30 340 230]{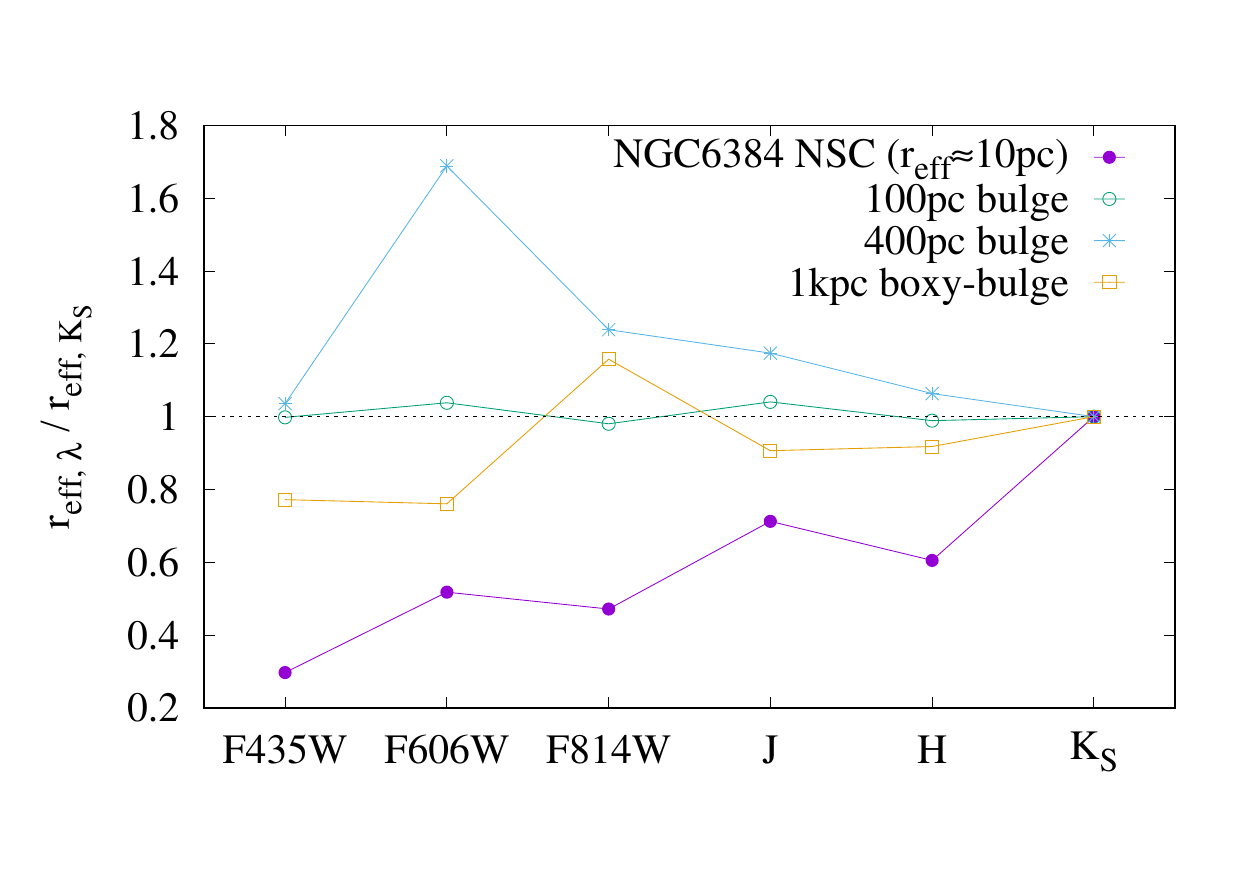}
\caption{Ratio between the effective radius in $K_S$-band to that in the other 
filters for each structural component.
}\label{fig:imfit_prop_table}
\end{figure}

We observe that the effective radius of NGC\,6384 NSC is larger at 
longer wavelengths. This is illustrated in Figure\,\ref{fig:imfit_prop_table}, alongside 
with the other fitted components, except for the large-scale exponential disk, which extent 
is well beyond our fitting area and thus might not be fully representative. Figure\,\ref{fig:imfit_prop_table} 
shows the ratio between the effective radius in the $K_S$-band, as a base, and that in 
the other filters. Especially for the NSC, we see that it becomes significantly more compact 
in $F435W$ (B-band) which suggests for a more centrally concentrated younger stellar 
population, as observed also in other studies \citep{Seth08b,Georgiev&Boeker14,Carson15}. 
On average, the $r_{\rm eff}$100\,pc bulge size is fairly invariant, while the 400\,pc bulge 
size is increasing with decreasing wavelength peaking in $F606W$. This filter also contains 
the $H\alpha$ line, which could be driving its very large size. Its very compact size in $F435W$ 
supports the presence of more centrally concentrated young population. The NGC\,6384 boxy-bulge 
also shows steadily decreasing effective radius with decreasing wavelength. This implies that 
it is mainly composed of old stellar population. As the main focus of this paper is on the 
NSC of NGC\,6384 would like to not extend our discussion to the bulge/pseudo-bulge topic.

The structural reanalysis of the HST WFPC\,2 $F606W$ image, which we presented in 
\cite{Georgiev&Boeker14}, showed that not accounting for the inner components led to an 
overestimate of the NSC effective radius by $\sim30\%$ and its magnitude by $\sim0.4$\,mag. 
Overall, this might not represent a drastic bias (roughly 0.4\,dex biased mass estimate), however, 
the ability to resolve and model all components in the central regions is essential for the correct 
mass and dynamical modelling of the NSC and its surroundings when future spectroscopy is 
included.

\subsection{NSC stellar population from the SED analysis}\label{Sect:Discussion_SED}
Using the six band photometry, we perform in \S\,\ref{Sect:SED} an SED fitting to the model 
magnitudes of the NSC (Fig.\,\ref{fig:NSC_SED}) as well as the to the SB-profile of NGC\,6384 
(\S\,\ref{fig:SED_map}). We find that the NIR and optical data helped to leverage the 
age-metallicity-extinction degeneracies and derive the effective NSC stellar population properties 
(cf. Table\,\ref{Table:SED}). As typically observed in NSCs most of the mass/light is 
contained in two populations of young and old age, the latter $\simeq\!90\%$ 
by mass \cite[e.g.][]{Seth06}, we find for the NGC\,6384 NSC an old component with an age of 
$10.8\pm1.3$\,Gyr and a stellar mass of ${\cal M}_{\rm\star,NSC,old}\!=\!3.5\times10^7M_\odot$. 
The a young component has an age of $230\pm140$\,Myr, which is $\sim8\%$ of the mass of 
the old population. The e-folding time for the old and young population is 2.3 and 0.14\,Gyr, 
respectively. The old and young stellar populations have slightly sub- and super-solar metallicity, 
which is in line with spectroscopic metallicity measurements of other extragalactic NSCs. However, 
because our SED fitting can report only for the total metal content (Z/H), therefore the metallicity 
values in Table\,\ref{Table:bestfits} are strong upper limits to the $Fe$-abundance. 

The radial map of the NSC surface mass density, age, metalicity and reddening 
(cf. Fig\,\ref{fig:SED_map}) follow trends consistent with the afore mentioned breakdown 
of the stellar population of the NSC. The extinction value of $E(B-V)\simeq1$\,mag, 
i.e. $A_V\simeq4$\,mag is expectedly, much lower than that toward the Galactic center, 
due to the lower inclination of NGC\,6384 nucleus, compared to the line of sight toward 
the MW NSC. The reddening values found here are in agreement with the range found for 
the nuclear regions of other galaxies from optical-NIR photometric analysis \cite[e.g.][]{Ganda09,Carson15}.

\subsection{Insufficient evidence for AGN activity}\label{Sect:Discussion_AGN}
The SED fitting resulted in negligible, $<\!1$\% fraction of the light ($\lesssim\!10^40$\,erg/s), 
to be coming from a weak AGN activity. This would be the upper limit on the 
possible AGN contribution to the NSC SED. We have only six SED points, which still 
leaves the possibility of a weak AGN component, however, its geometry has to have 
a more extreme configuration than the one found here (cf \S\,\ref{Sect:SED}). Any 
present degeneracies should be reflected in the uncertainty ranges obtained from the 
posteriors by {\sc CIGALE}. Also, if the AGN fraction is higher, instead of a contribution 
from younger stellar population, this would lead to a steeper SED continuum shape in 
the NIR (redder colour), and will require lower reddening. However, this will be inconsistent 
with the overall optical-NIR SED shape, as investigated observationally by SED studies 
at pc scales to identify AGN contributions \cite[e.g.][]{Prieto10}. This highlights the power 
of combining optical and NIR that covers one of the most important wavelength ranges 
for minimising various of the afore mentioned degeneracies. Therefore, the current data 
can not support an active MBH in the NSC of NGC\,6384. This, however, does not exclude 
the presence of a MBH, because it might not be in an active state, just like the one in 
the MW and M\,31. A follow up high spatial and spectral resolution NIR spectroscopy 
is needed to measure the NSC and its central velocity dispersion to assess the presence 
of a MBH. The lack of clear AGN component in our SED fitting and the indications for 
very centrally concentrated young stellar population as discussed earlier 
(cf. Fig.\,\ref{fig:imfit_prop_table}) suggests that the LINER emission of the nucleus of 
NGC\,6384 measured in low-spatial resolution ($2.5\arcsec$ spectra) is likely due to the 
young star formation activity in its central regions rather than an AGN activity.

\subsection{Implications for the formation of NGC\,6384 NSC}\label{Sect:Discussion_NSCform}
Finally, the evidences from our measurements of the NGC\,6384 NSC having an effective 
radius as extended as $r_{\rm eff}\simeq10$\,pc and varying with wavelength, two circum nuclear Sersi\'c components 
of 100 and 400\,pc scales and low $n-$ indices consistent with being nuclear disks, the 
non negligible by mass young stellar component and the clear detection of disk-like dust 
lane spiralling into the cluster shares features of the two main scenarios of NSC formation: 
$i)$ cluster merging, which leads to larger effective radii and possible formation of nuclear 
disks (see \S\,\ref{Sect:Discussion_Multiponents}) $ii)$ repetitive gas accretion which leads 
to the presence of young stellar population (see \S\,\ref{Sect:Discussion_SED}). 
This finding for NGC\,6384 NSC adds to the growing body of evidence in the literature that 
supports not a single formation channel during the build up of galactic nuclei, but a complex 
and individual mixture of both. A larger, representative sample of galactic nuclei analysed 
in such manner can allow for a more systematic and statistically meaningful conclusions 
about the build up of galactic nuclei depending on galaxy mass, type, environment, and 
whether the nucleus is co-inhabited by a MBH.

\section{Summary}\label{Sect:Summary}

Here we presented the analysis of the first science data taken during the commissioning 
of the ARGOS system in 2015-05-01 and 02. It provides adaptive optics correction of the 
ground atmospheric layer at the LBT. The target for this commissioning run, NGC\,6384 
was mainly selected because it has a suitable on-axis AO reference star and a large number 
of MW stars to measure system performance. The fact that this galaxy was at a large distance 
of 20.7\,Mpc allowed us to also demonstrate that star cluster science can be conducted 
successfully, where ground based seeing limited observations lack the needed spatial 
resolution to study such compact stellar systems. 

For the proper analysis of the NIR images with the LUCI\,2 camera, which 
suffers from persistence and non-linearity effects, we created pixel-to-pixel maps that 
we used to preprocess and correct the raw images (details in \S\,\ref{Sect:App.NonLin}). 
Following standard data reduction steps (\S\,\ref{Sect:DataReduction.LUCI2}) and calibration, 
we were able to achieve excellent image registration resulting in a sharp and stable PSF 
of 0.25\arcsec\ over the entire $4\arcmin\times4\arcmin$ LUCI\,2 field of view. The PSF 
size only increases by $\lesssim\!25\%$ out to the detector edges (\S\,\ref{Sect:App.PSF}). 
This superb spatial resolution enables to resolve the star cluster candidates and bring down 
contamination from background galaxies to a minimum. In a forthcoming paper we will 
present the detailed analysis of the star cluster system of NGC\,6384 (cf \S\,\ref{Sect:Phot}), 
while in this paper we mainly focus our analysis to the central $3.6\!\times\!3.6$\,kpc 
($\sim\!36\arcsec\!\times\!36\arcsec)$ overlap region between our NIR LBT and archival 
optical HST data. The high spatial NIR imaging is crucial for breaking age-metallicity-reddening 
degeneracies when combined with the optical HST data. We reprocessed the HST/ACS 
and WFPC\,2 data to the same plate scale resolution of 0.05\arcsec/pix (cf. 
\S\,\ref{Sect:DataReduction.HST}).

We performed a 2D MCMC image fitting with {\sc imfit} \citep{Erwin15} by using a PSF 
model built from stars in the image for both the LUCI\,2 and HST images, as well as a 
{\sc TinyTim} \citep{Krist&Hook11} PSF model drizzled in identical manner as the science 
images. We also used a pixel mask of the central obscuring disk-like dust lane (\S\,\ref{Sect:Struct:NSCGal}, 
Fig.\,\ref{fig:ellipse_rsd}). 

Our main findings can be summarizes as:\\
\indent$\bullet$ We uncover the presence of two inner Sersi\'c (low $n-$index, disk?) 
components (\S\,\ref{Sect:Struct:NSCGal}, Fig.\,\ref{fig:SB} and \S\,\ref{Sect:Discussion_Multiponents}) 
with effective radii of $\sim\!100$ and 400\,pc in which the NSC with 
$r_{\rm eff}\simeq10$\,pc is embedded. NGC\,6384 has a 
large scale boxy bulge ($\sim1$\,kpc) and disk ($\sim6$\,kpc).

\indent$\bullet$ The effective radius of NGC\,6384 NSC increases with wavelength (cf. Fig.\,\ref{fig:imfit_prop_table}, 
which suggests for a more centrally concentrated younger stellar population (see \S\,\ref{Sect:Discussion_reff_variat}).

\indent$\bullet$ The NSC effective radius is smaller in size by $\sim30\%$ and magnitude/mass 
by $\sim0.4$\,mag/dex compared to \cite{Georgiev&Boeker14} due to the unaccounted inner 
Sersi\'c components. This highlights the need to resolve and model well the central regions for 
the correct mass and dynamical modelling of the NSC when follow up spectroscopy is included.

\indent$\bullet$ The NSC SED is described by an old population ($10.8\pm1.3$\,Gyr) with a 
stellar mass of ${\cal M}_{\rm\star,NSC,old}\!=\!3.5\times10^7M_\odot$, and a young population 
of $230\pm140$\,Myr, which is $\sim8\%$ of its mass. Both populations have slightly sub- 
and super-solar metallicity, respectively (cf. Table\,\ref{Table:SED},\S\,\ref{Sect:SED} and 
\S\,\ref{Sect:Discussion_SED}).

\indent$\bullet$ We obtained the SED radial surface mass density, age, metalicity and reddening 
(cf. Fig\,\ref{fig:SED_map}, \S\,\ref{Sect:SED}, \S\,\ref{Sect:Discussion_SED})

\indent$\bullet$ We find negligible, $<\!1$\% fraction of the light, to be coming from 
a weak AGN (\S\,\ref{Sect:Discussion_AGN}). If a MBH in the NSC of NGC\,6384 is present, 
then it might not be accreting, similarly to those in the MW and M\,31.

\indent$\bullet$ All structural and stellar population evidences suggest that the NGC\,6384 NSC 
for a formation contributed from the two main scenarios of NSC formation: cluster merging and 
repetitive gas accretion.

\section*{Acknowledgements}
{\small
We thank the anonymous referee for their constructive and useful comments that helped to improve parts of the discussion in the paper. 
Based on observations collected at the Large Binocular Telescope (LBT). The LBT is an international collaboration among institutions in the United States, Italy and Germany. LBT Corporation partners are: The University of Arizona on be- half of the Arizona Board of Regents; Istituto Nazionale di Astrofisica, Italy; LBT Beteiligungsgesellschaft, Germany, representing the Max-Planck Society, The Leibniz Institute for Astrophysics Potsdam, and Heidelberg University; The Ohio State University, and The Research Corporation, on behalf of The University of Notre Dame, University of Min- nesota and University of Virginia.
} % Closing \small

%%%%%%%%%%%%%%%%%%%%%%%%%%%%%%%%%%%%%%%%%%%%%%%%%%



\bibliographystyle{mnras} 
\bibliography{references}

%%%%%%%%%%%%%%%%%%%%%%%%%%%%%%%%%%%%%%%%%%%%%%%%%%

%%%%%%%%%%%%%%%%% APPENDICES %%%%%%%%%%%%%%%%%%%%%

\appendix
\section{LUCI\,2 linearity and persistence maps}\label{Sect:App.NonLin}

The first linearity and persistence analysis of the LUCI\,2 N\,3.75 detector was 
performed by David Thompson (at LBTO) that is available at the LUCI\,2 
webpage\footnote{\href{http://scienceops.lbto.org/sciops\_cookbook/luci2-vs-luci1/\#L2vL1\_Nonlinearity}{http://scienceops.lbto.org/sciops\_cookbook/luci2-vs-luci1/\#L2vL1\_Nonlinearity}}. 
This linearity correction needs to be applied to the counts of all raw data prior 
any data reduction. The coefficients for this correction are estimated for the 
detector average pixel values, which can be a good overall correction. However, 
often (if not exclusively), science measurements are made on a small pixel area 
of the detector, which could (significantly) deviate from the detector average 
linearity coefficients and compromise scientific flux and position measurements. 
Therefore, we performed a new pixel-by-pixel linearity, and not quantified so far, 
persistence analysis of the LUCI\,2 N\,3.75 camera. For that we used linearity 
data taken and kindly provided to us by David Thompson. The data consists of 
a sequence of images with increasing exposure time until detector saturation is 
reached. Two consecutive frames are taken at the same exposure time to allow 
to analyse detector persistence. For the analysis, we extract the count value 
of every pixel and exposure and fit curves to the 1st and 2nd exposures as a 
function exposure time. 

\begin{figure}
\includegraphics[width=.48\textwidth]{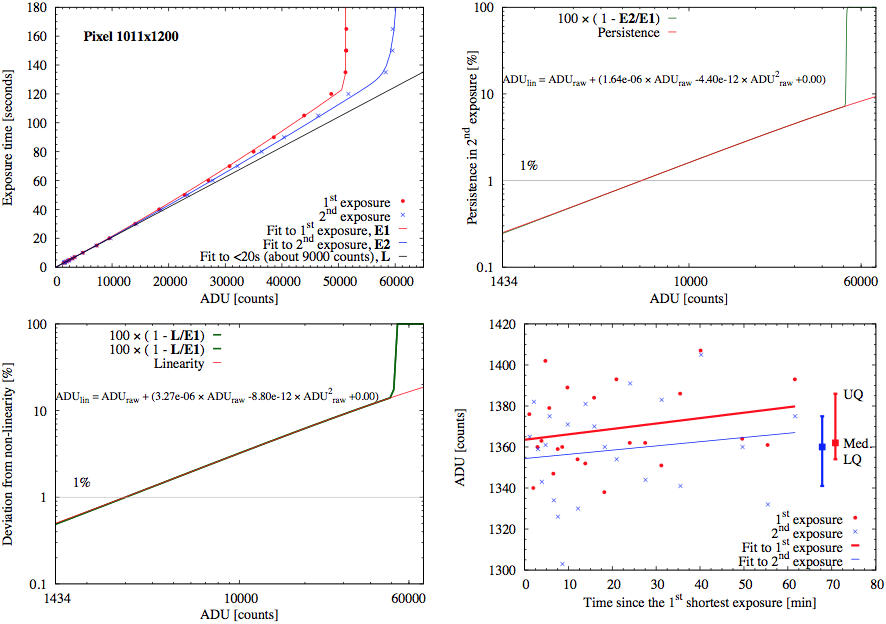}
\includegraphics[width=.48\textwidth]{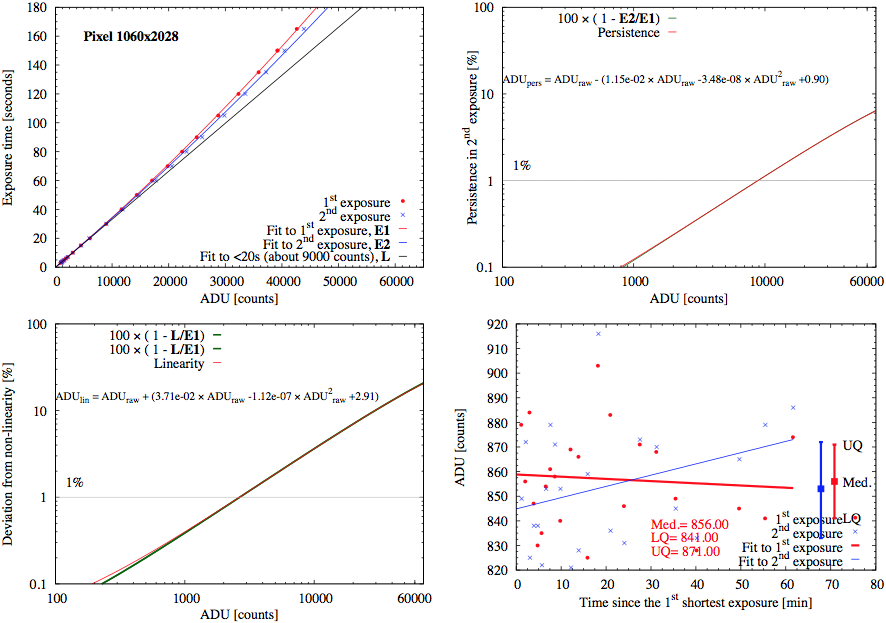}
\caption{An example for the pixel-by-pixel linearity and persistence solutions for 
the LUCI\,2 detector. {\bf Top four} and {\bf bottom four} panels show an example of two pixels at different 
location on the detector that show different linearity and persistence curves. For 
each group of four panels, from {\bf top-left} to {\bf bottom-right} clockwise, are 
shown the exposure time as a function of pixel value (shown this way for fitting purposes), 
the linearity, the persistence and the noise statistics of that pixels as a function 
of the time during which the linearity data was taken. Lines with solid circles and 
crosses in {\bf top-left} panel show the 1st and second exposure taken in a sequence 
until saturation is reached and their fitted functions, as indicated in the panel legend. 
{\bf Top-right} and {\bf bottom-left} panels show the ratio of the fitted functions 
and their coefficients are shown with label in the panels. {\bf Bottom-right} panels 
show the pixel value of the 1st and 2nd exposure in the sequence 
}\label{fig:linearity_fit}
\end{figure}

We confirm results from earlier analysis that departures 
from linearity $\gtrsim\!5\%$ occur at around 9000 counts. Therefore, to set the 
linearity reference line we perform a least squares fit of a linear form to the 
count values of the 1st exposure smaller than 9000 counts. The ratio between the 
curves of the 1st exposure and that of the linear region provides the linearity 
correction relation. The ratio between the fitted curves for the 1st and 2nd 
exposure sets as a function of the count rate provides the relation for the 
persistence correction. Each of these fits and ratios, for the persistence and 
linearity, are shown in Figure\,\ref{fig:linearity_fit}. To demonstrate the 
ranging behaviour of different pixels, in the figure we show the fits for two 
pixels at different position on the detector. Each pixel is represented by a four 
panel figure showing the fits and linearity and persistence correction equations 
and their coefficients, as described in the figure legend. In the bottom-right panel 
we show the pixel noise statistics and the fit through it as a function of the duration 
time of the entire calibration sequence. The latter plot shows that there are no 
strong trends as a function of time, which might be expected to accumulate due to 
the persistence.

\begin{figure}
\includegraphics[width=.5\textwidth]{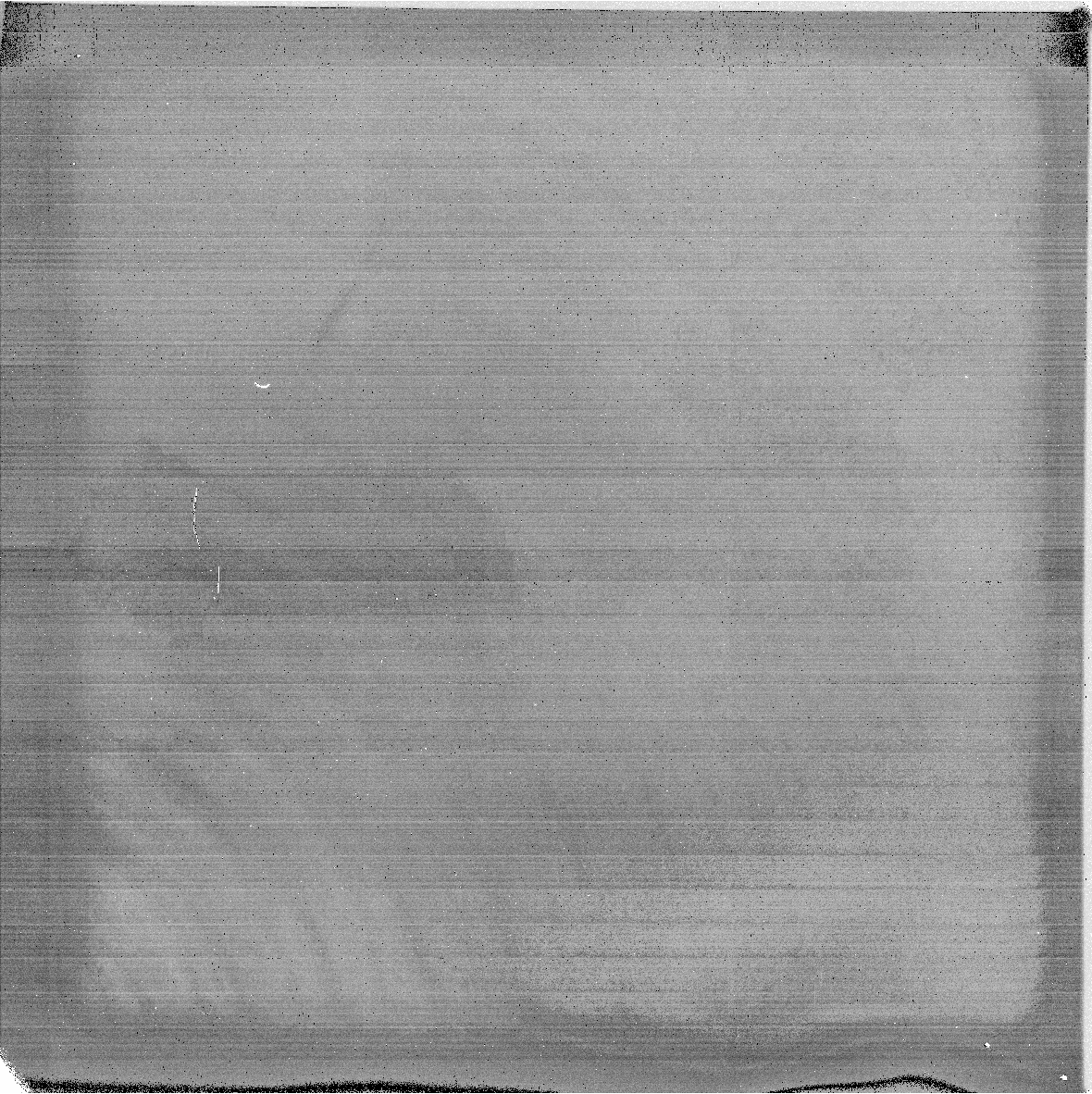}
\caption{An example of LUCI\,2 N\,3.75 detector linearity map of the linear coefficient. 
Such maps are available for all linearity and persistence coefficients. The maps give 
the (zero point, linear and quadratic term) coefficient values derived for each pixels, 
as shown by the examples in Fig.\,\ref{fig:linearity_fit}
}\label{fig:linearity_maps}
\end{figure}

\begin{figure}
\includegraphics[width=0.498\textwidth]{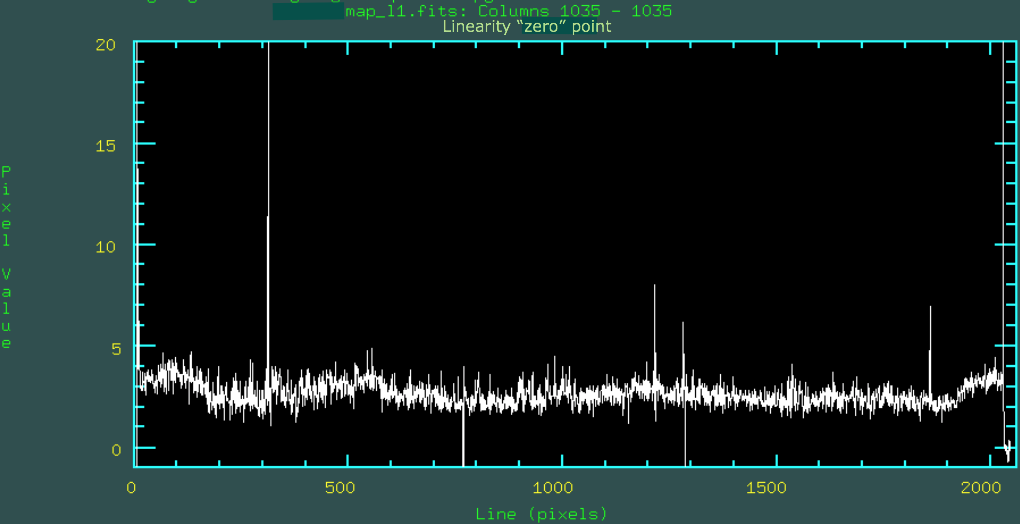}
\includegraphics[width=0.498\textwidth]{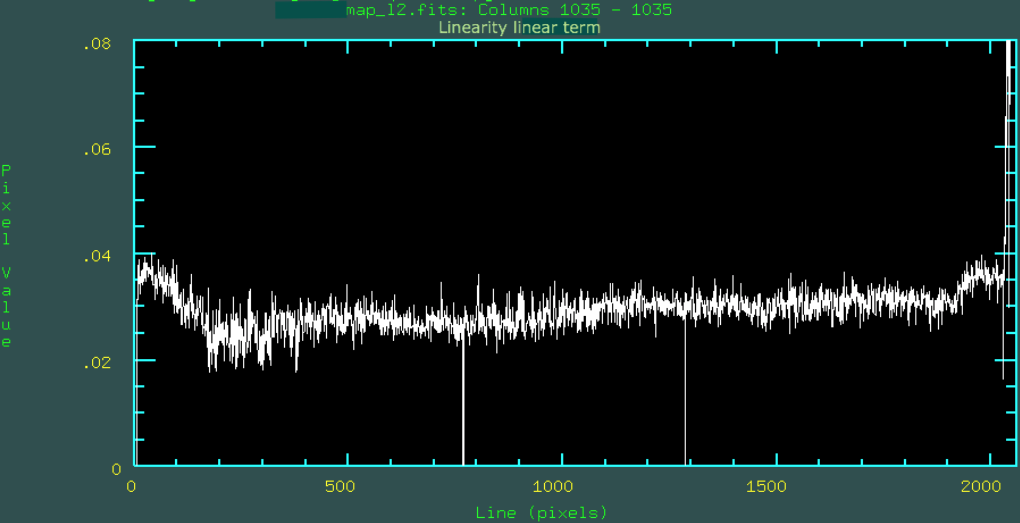}
\includegraphics[width=0.498\textwidth]{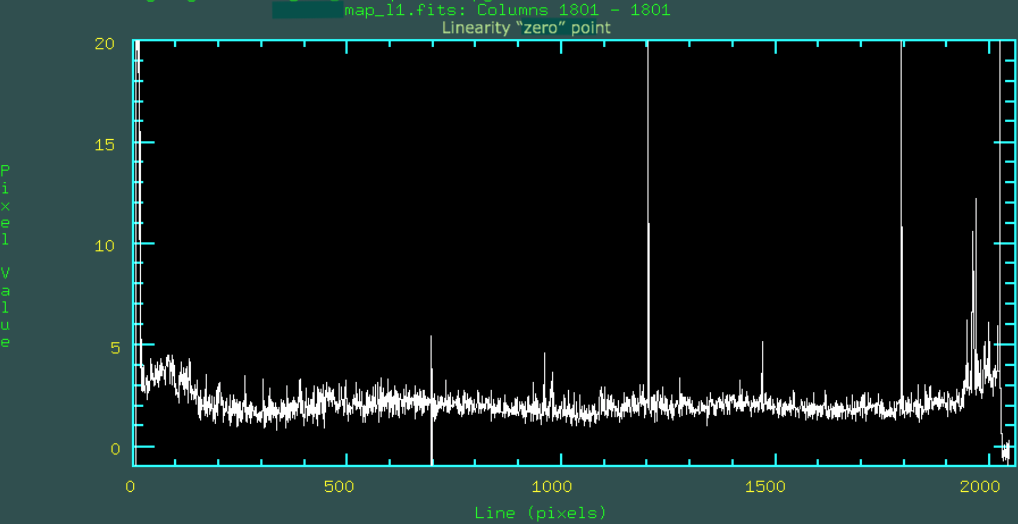}
\includegraphics[width=0.498\textwidth]{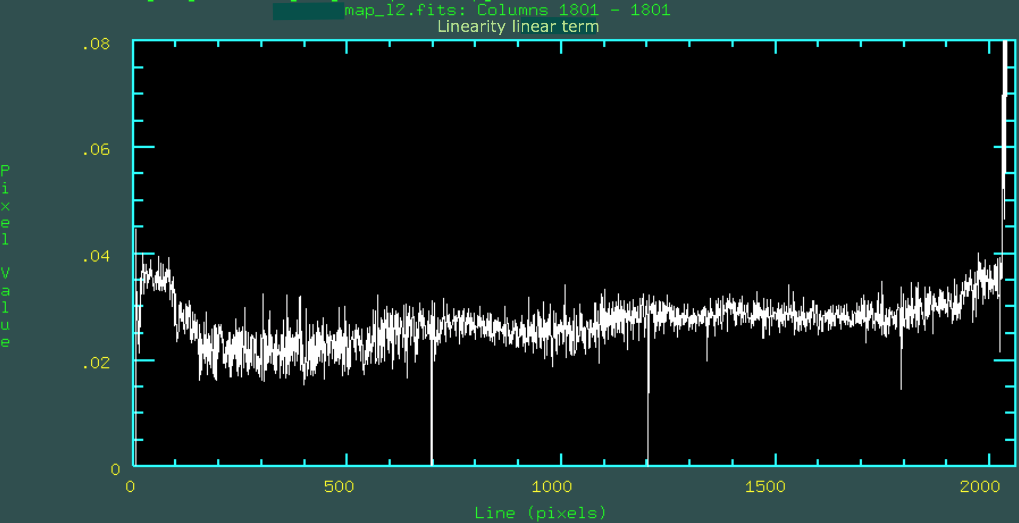}
\caption{An example of linearity zero point ({\bf left}) and linear coefficient 
({\bf right}) values at two different detector X-pixel positions ({\bf top} and 
{\bf bottom} row) cutting through all detector Y-pixel locations. Cold and hot 
spikes in the plots indicate cold, hot or bad pixels.
}\label{fig:linearity_cuts}
\end{figure}

The zero point, linear and quadratic coefficients from the linearity and persistence 
fits, performed to each pixel of the detector, are stored as six fits image maps. An 
example of one such map is presented in Figure\,\ref{fig:linearity_maps}, which shows 
the map of the coefficient of the linear term. With this map one can appreciate the 
large and small scale variation of this coefficient. A number of detector features 
are visible such as groups of bad, hot or cold pixels and a horizontal stripes pattern 
from the detector electronics which repeats every 64th column. To appreciate the 
pixel-to-pixel variation in Figure\,\ref{fig:linearity_cuts} we show cuts through 
the zero point and linear term maps at two different x-pixel locations, 1035 and 1801. 
As already visible in Fig.\,\ref{fig:linearity_maps}, the large and small scale trend 
and noise, is well visible in Fig.\,\ref{fig:linearity_cuts}. Positive and negative 
spikes show hot and cold or bad pixels.

\section{ARGOS PSF performance}\label{Sect:App.PSF}

\begin{figure*}
\includegraphics[width=1.05\textwidth]{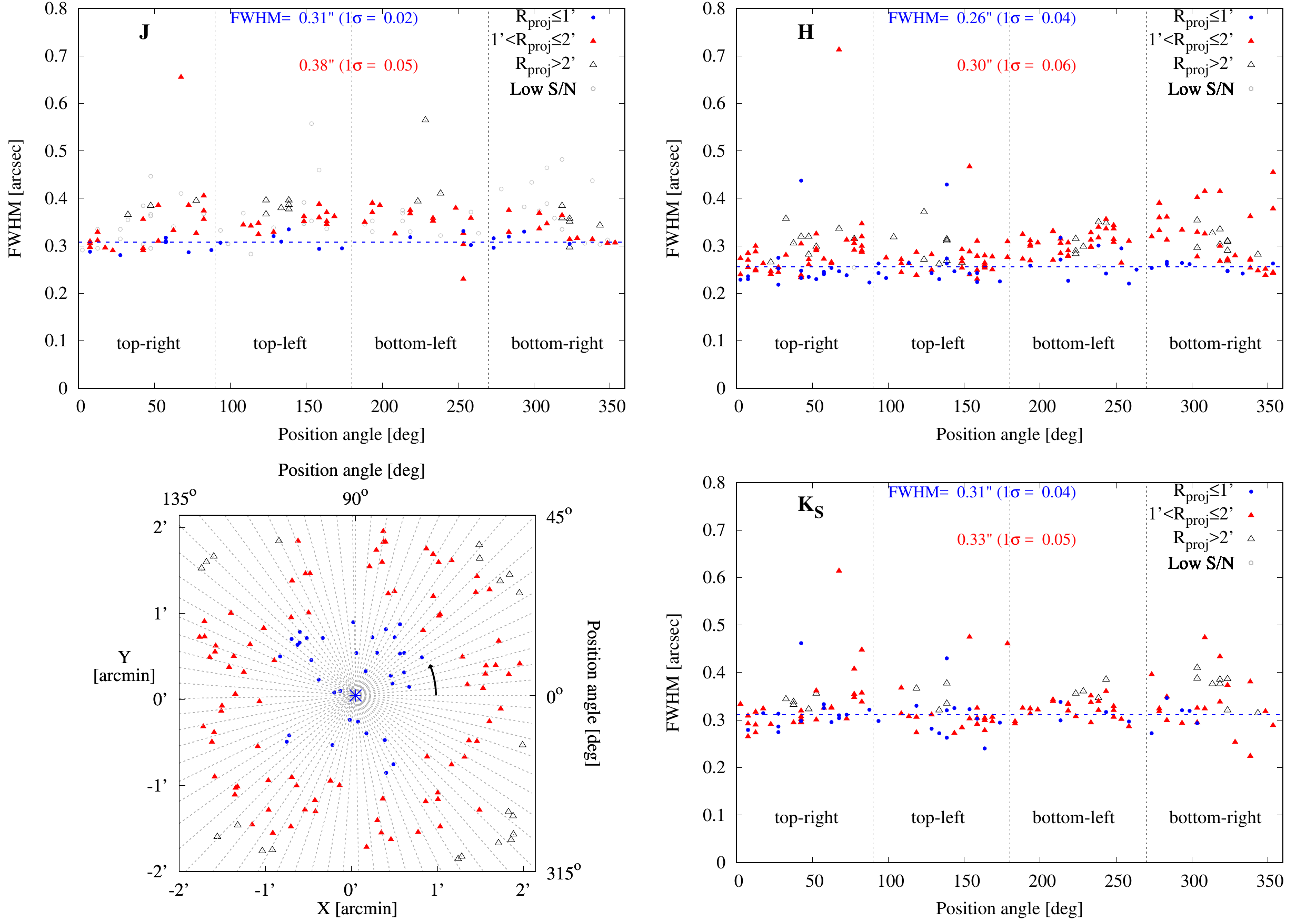}
\caption{ARGOS commissioning data analysis of the angular and radial PSF$_{\rm FWHM}$ 
variation. The {\bf top two} and {\bf bottom-right} panels show the azimuthal variation 
of the FWHM in the three filters as labelled. Every symbol is a source in the LUCI\,2 
image with a $S/N\!>\!15$, at a given position angle (PA) and at a radial distance from 
the NGS (image centre) of $r\!\leq\!1'$, $1\!\leq\!r\!\leq\!2'$ and $r\!>\!2'$, as indicated 
in the figure legend. With a horizontal dashed line is shown the best least squares fit 
of the FWHM through the data within $r\!\leq\!1'$, where its value and $1\sigma$ dispersion 
is shown with labels, as well as that for data within $1\!\leq\!r\!\leq\!2'$. In the {\bf bottom-left} 
panel curved arrow starting at $0^\circ$ shows the PA and its counter clockwise direction 
of rotation. To guide the eye, with dashed lines are also shown each $5^\circ$ of PA.
}\label{fig:fwhm_spatial}
\end{figure*}

To correct the ground layer turbulence, ARGOS uses a constellation of three lasers per 
telescope evenly situated on a circle with radius of 2', which also corrects for the 
strong anisoplanatism 
as a function of distance from the natural guide star (NGS). For this particular first 
science data observation, the NGS was located at the image center, therefore, it is expected 
that the effects on the image PSF (elongation, orientation) should be symmetric. We use 
the measurements from our PSF modelling and photometry described in detail in \S\,\ref{Sect:Phot}. 
The field of NGC\,6384 contains a large number ($>\!100)$ of foreground Galactic stars.
To analyse the angular and radial variation of the ARGOS corrected image PSF, in 
Figure\,\ref{fig:fwhm_spatial} we show the FWHM of 
the PSF (PSF$_{\rm FWHM}$) as a function of the position angle (PA) and the observed $JHKs$ 
filters in the top row and bottom-right panels. With different symbols and colours we show 
the three different radial bins, as indicated in the figure legend. The bottom left panel 
of Fig.\,\ref{fig:fwhm_spatial} illustrates the direction of the position angle shown in 
the other panels and the spatial position of the high-$S/N$ sources on the detector. We 
fitted the FWHM of the sources within 1' and between 1' and 2' distance from the NGS, which 
are shown as labels for the respective filter in the different panels. Due to the relatively 
large LUCI\,2 field of view for adaptive optics correction, it is expected that some anisoplanatism 
could still be present as a function of the distance from the NGS to the LGSs, which are 
situated on a circle of radius 2'. We see from Figure\,\ref{fig:fwhm_spatial} that the FWHM 
fits to sources within 1' and between 1' and 2' distance from the NGS differ by up to 
$\sim\!20\%$. Sources outside the 2' radius in the image corners are also seen to be slightly 
more extended (often with the highest FWHM value), however, due to their relatively low number, 
we can not fully quantify the FWHM degradation outside the LGSs circle from this data alone. 
Overall, for this first commissioning science data and sky quality during the observations, 
the achieved FWHM correction provided by ARGOS in the NIR is around 0.25''-0.3'', which is 
also very stable across the detector to within $\sim\!20\%$.

\begin{figure*}
\includegraphics[width=1.05\textwidth]{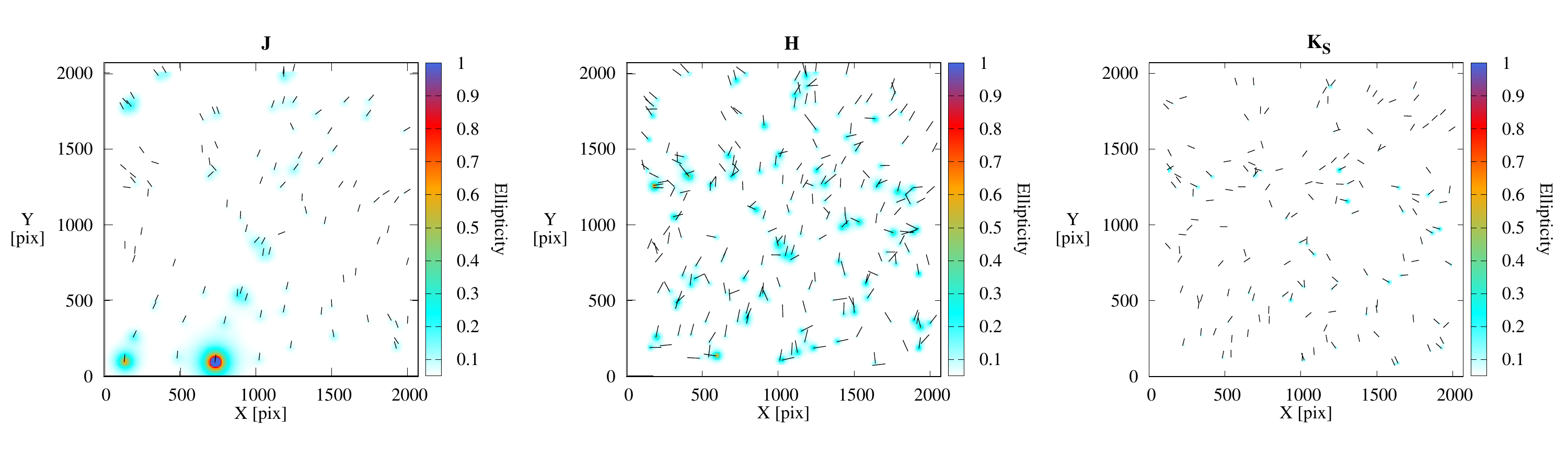}
\caption{PSF ellipticity and orientation analysis. Vectors show the direction of elongation 
of the PSF for high-$S/N$ sources, which are also colour coded by their PSF ellipticity. The 
three panels show the data for the three different filters.
}\label{fig:fwhm_ell}
\end{figure*}

We examine the elongation and orientation of the PSF$_{\rm FWHM}$ in Figure\,\ref{fig:fwhm_ell}. 
The vectors in the figure show the direction of orientation of the PSF for high-$S/N$ sources, 
which are also colour coded according to their ellipticity. Measured PSF orientation makes 
sense only for sources with ellipticity larger than about 0.05 (axis ratio $\gtrsim5\%$). We can indeed see that most of the sources toward 
the detector corners, more apparent in $J$-band (Fig.\,\ref{fig:fwhm_ell} left panel), show 
more radially aligned orientation.

\begin{figure}
\includegraphics[width=0.48\textwidth]{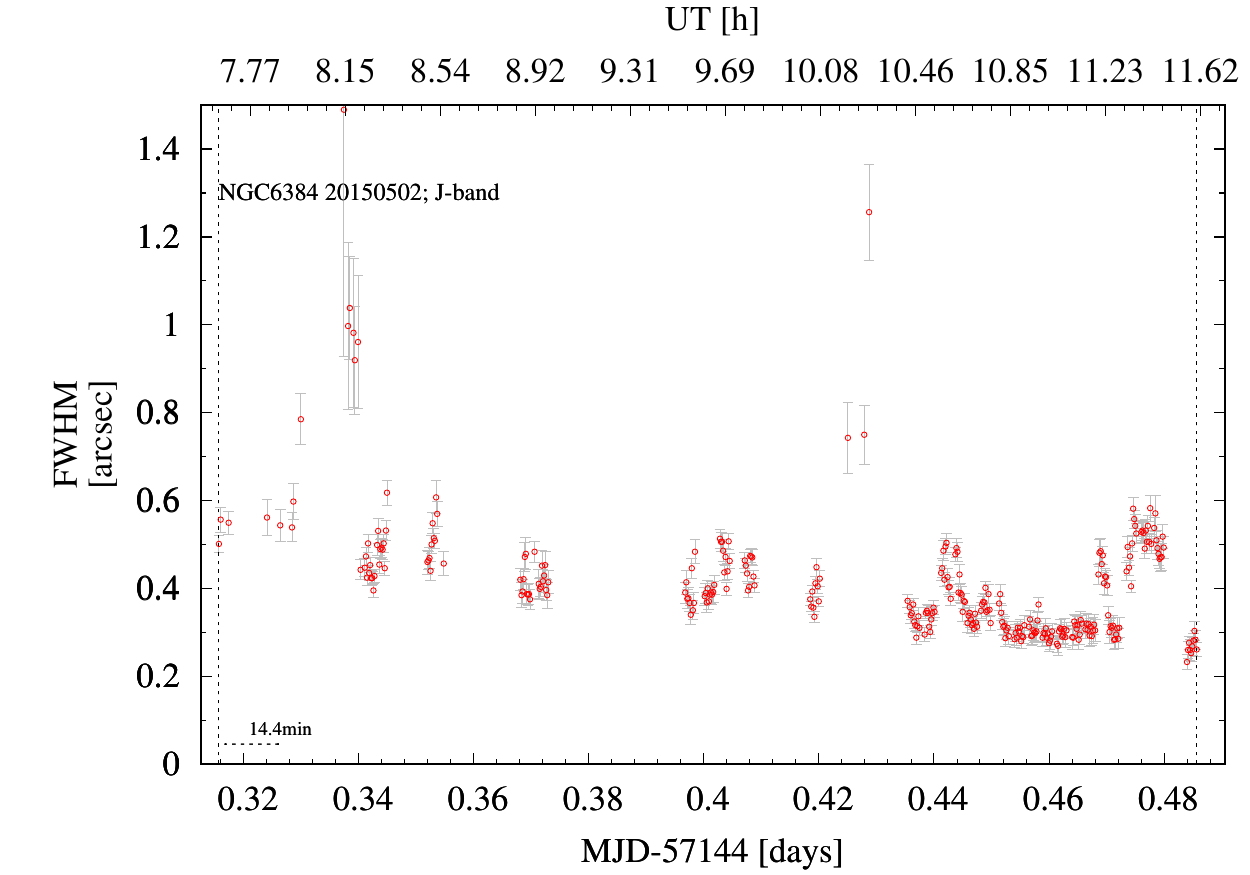}
\includegraphics[width=0.48\textwidth]{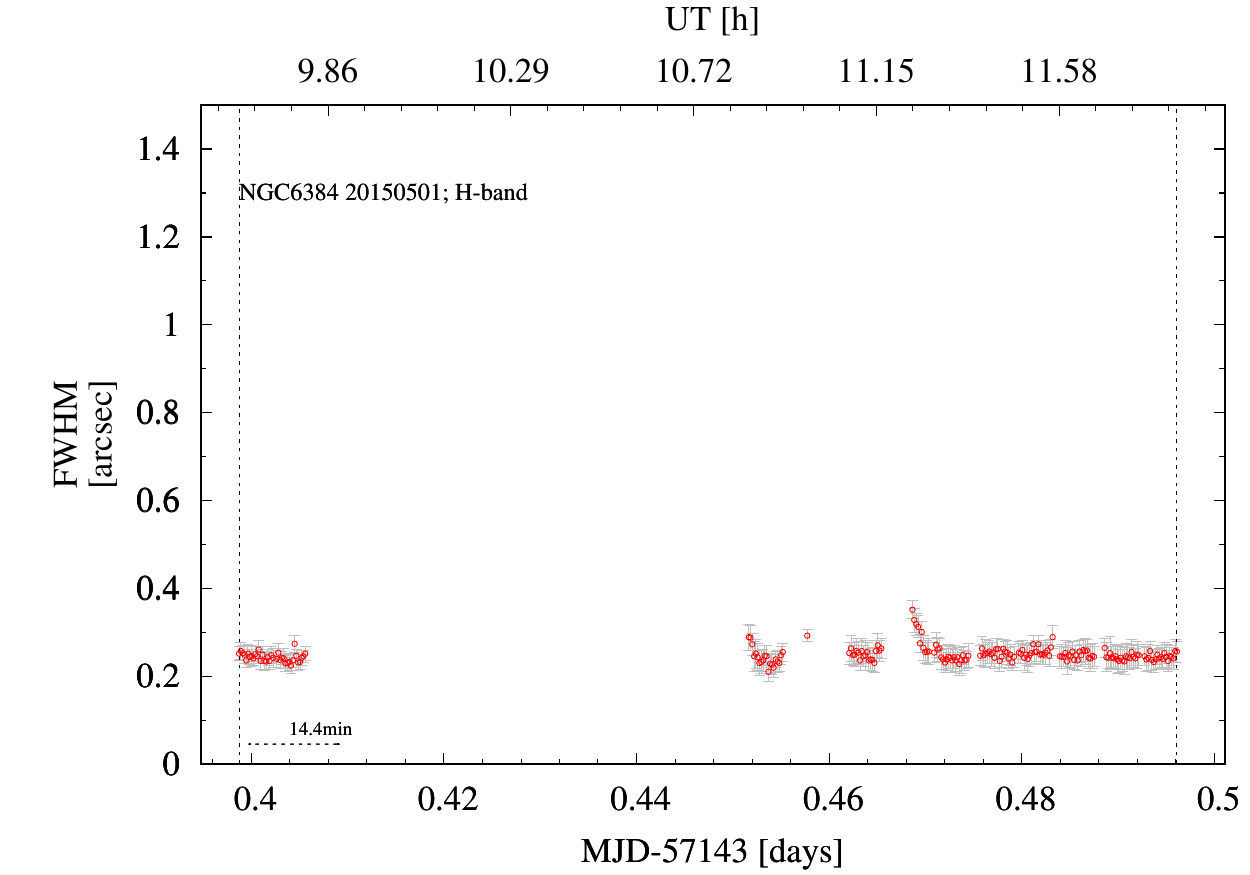}
\includegraphics[width=0.48\textwidth]{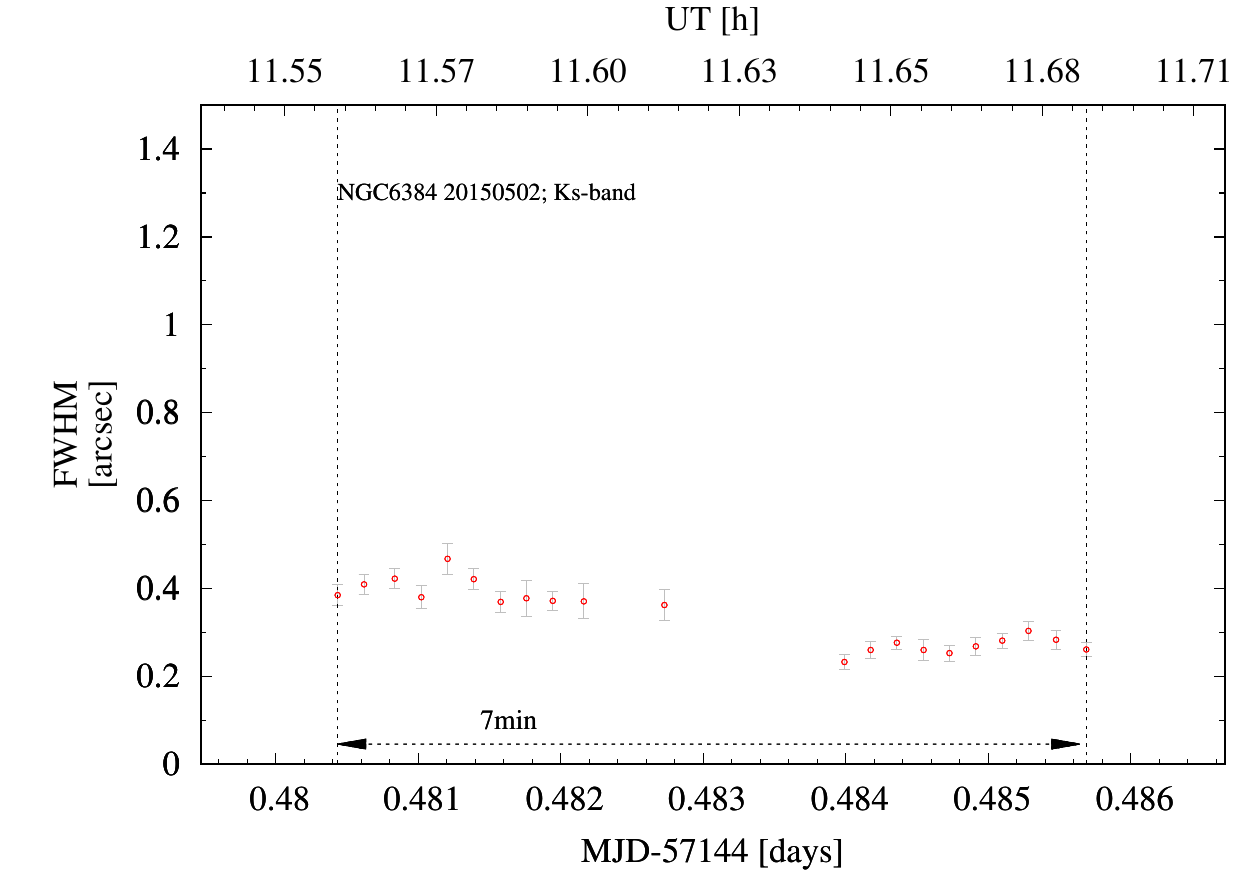}
\caption{ARGOS commissioning data analysis of the temporal PSF$_{\rm FWHM}$ variation. 
From top to bottom is shown the variation for the three filters, $J,$ $H$ and $Ks$, as a 
function of time (in MJD [days] on the x-axis and UT [h] on the x2-axis).
}\label{fig:fwhm_time}
\end{figure}

The absolute value of the corrected PSF$_{\rm FWHM}$ is also a function of the natural seeing 
and sky conditions (e.g. passing high altitude clouds, variable direction and strength of wind), 
which obviously can change within the night or from one night to the other. In addition, for 
such a first commissioning run the ARGOS system performance is not yet optimal and improvement 
was in progress. To analyse these temporal variations of the image FWHM in Figure\,\ref{fig:fwhm_time} 
we show the image FWHM for the three filters ($JHKs$ top to bottom) as a function of the time 
when an exposure was taken. We have limited here our analysis only to the datasets which were 
(partly) used to select the sharpest images for the scientific analysis of the data as described 
in the main text of this paper. Therefore, we are not showing here the FWHM variation during 
various phases of the commissioning testing, which is presented in a dedicated ARGOS commissioning 
paper by \cite{Rabien18}. As it can be seen, the best and most stable 
image quality (FWHM) was achieved on the night of 2015-05-01 in the $H$-filter (Fig.\,\ref{fig:fwhm_time} 
middle panel). The following night of 2015-05-02 is showing a larger FWHM variation, but improving 
as a function of time (compare Fig.\,\ref{fig:fwhm_time} top and bottom panels). This large 
scale FWHM improvement (besides its absolute value due to the difference in the $J$ and $K_S$ 
PSF$_{\rm FWHM}$) is attributed to the improving sky conditions (decreasing wind, improving 
natural seeing). The variations on small temporal scales are due to broken laser loops, which 
are either caused by commissioning tests or often by a system pause due to satellite or an 
airplane passing overhead. The latter two requires lasers shut down and opening the loops. 
Images taken during that phase often show the natural seeing value in the respective filter. 
This can be seen in the top panel of Figure\,\ref{fig:fwhm_time} in the $J$-band, where loops 
were broken/paused about three times at around 8.15, 10.2 and 11.4 UT time. Apparently, the  
night quality was improving and the open loops FWHM value is decreasing. There were few filter 
changes between $J$ and $K_S$ at between 11.5 and 11.62 UT time, but most of the $K_S$-band 
observations (Fig.\,\ref{fig:fwhm_time} bottom) followed the $J$-band data. The combination 
between improving sky quality and sharper PSF in the $K_S$ provided a nearly diffraction 
limited sampling of the PSF of $0\farcs24\!=\!2$\,pix, as seen in Figure\,\ref{fig:fwhm_time} 
bottom panel and from the fits in Figure\,\ref{fig:fwhm_spatial} bottom-right panel.

Overall, in this section we showed that ARGOS provides LBT with spatially and temporary stable 
PSF$_{\rm FWHM}$ in the NIR over $4'\!\times\!4'$ field of view, which is unique for a ground 
based observatory.

\section{MCMC exploration of model parameters and their uncertainties}\label{Sect:App.imfit_mcmc}

\begin{figure*}
\subfloat[][]{\includegraphics[width=1\textwidth]{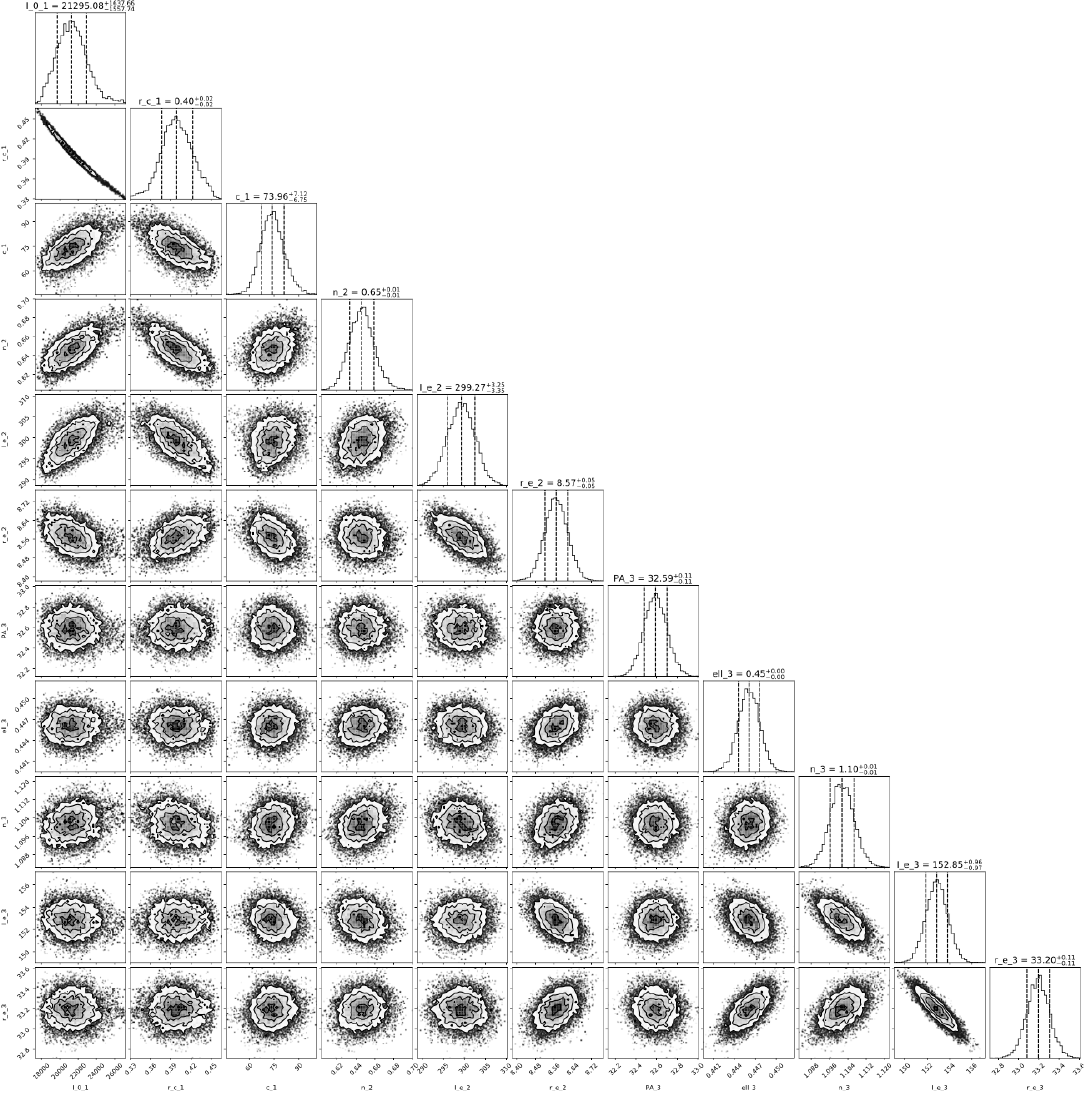}}
\caption{Corner plot of the fitted structural parameters on the $K_S$-band image obtained using 
the {\sc imfit} MCMC module. The best parameter values and their uncertainties were estimated 
using the median and their $1\sigma$ values indicated with vertical dashed lines. This plot 
shows the parameter values for the inner 3 components: King, Inner and Outer Sersi\'c components.
}\label{fig:imfit_mcmc}
\end{figure*}

\begin{figure*}
\ContinuedFloat
\subfloat[][]{\includegraphics[width=1\textwidth]{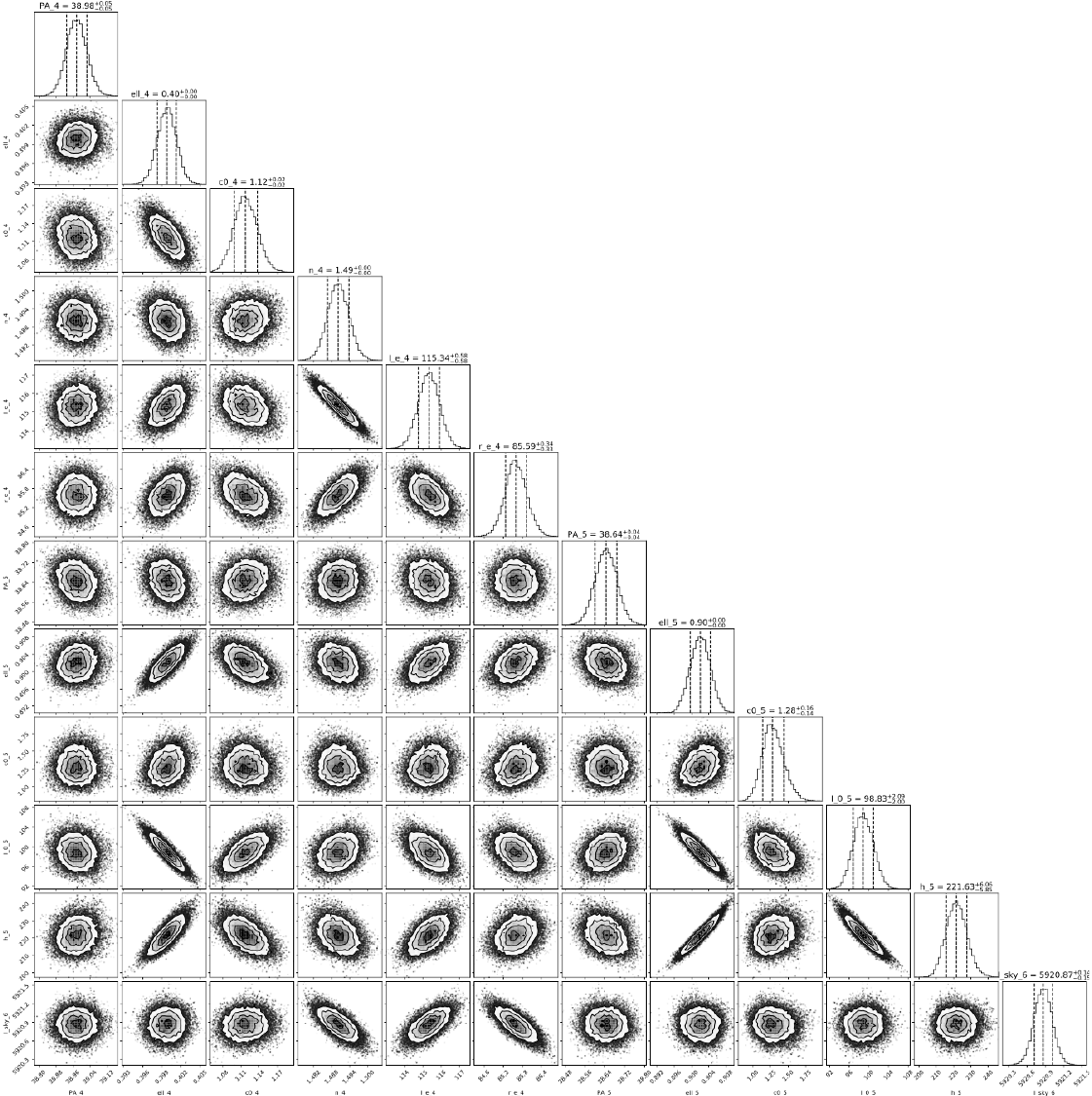}}
\caption{Cont'd. Parameter values for the Generalized Sersi\'c and Exponential profiles.
}\label{fig:imfit_mcmc}
\end{figure*}

As described in \S\,\ref{Sect:Struct:NSCGal}, we tested a wide variety, and number, of fitting 
functions that best describe the nuclear cluster and inner regions of NGC\,6384. A quick analysis  
was performed with $\chi^2$ minimisation, while the detailed parameter and their uncertainty 
exploration we conducted with the {\sc imfit} MCMC module. Here we show as an example of this 
analysis for the $K_S$-band and how we obtained the structural parameters and uncertainties 
for the NSC and NGC\,6384 inner $3.6\!\times\!3.6$\,kpc ($\sim\!36\arcsec\!\times\!36\arcsec)$. 
Figure\,\ref{fig:imfit_mcmc} shows the corner plot created with the corner module \citep{Foreman-Mackey16} 
of the Astropy channel \citep{Astropy13,Astropy18}. We explored in total 35 model parameters 
shared between the five functions - a modified King \citep{Elson99,Ch.Peng10} which for $\alpha\!=\!2$ 
reduces to the original \cite{King62}; two S\'ersic and two generalized elliptical 2D S\'ersic 
and Exponential profiles. Their functional forms are given in the 
\href{http://www.mpe.mpg.de/~erwin/resources/imfit/imfit\_howto.pdf}{{\sc imfit} manual}\footnote{\href{http://www.mpe.mpg.de/~erwin/resources/imfit/imfit\_howto.pdf}{http://www.mpe.mpg.de/$\sim$erwin/resources/imfit/imfit\_howto.pdf}}. 
As it can be seen from Figure\,\ref{fig:imfit_mcmc}, we obtained good convergence and uncertainty 
estimates for all parameters. 

\section{$r_{\rm eff}$ related to King model parameters}\label{Sect:App.r_eff_King}

\begin{figure}
\includegraphics[width=1.02\linewidth,trim=40 180 60 100,clip]{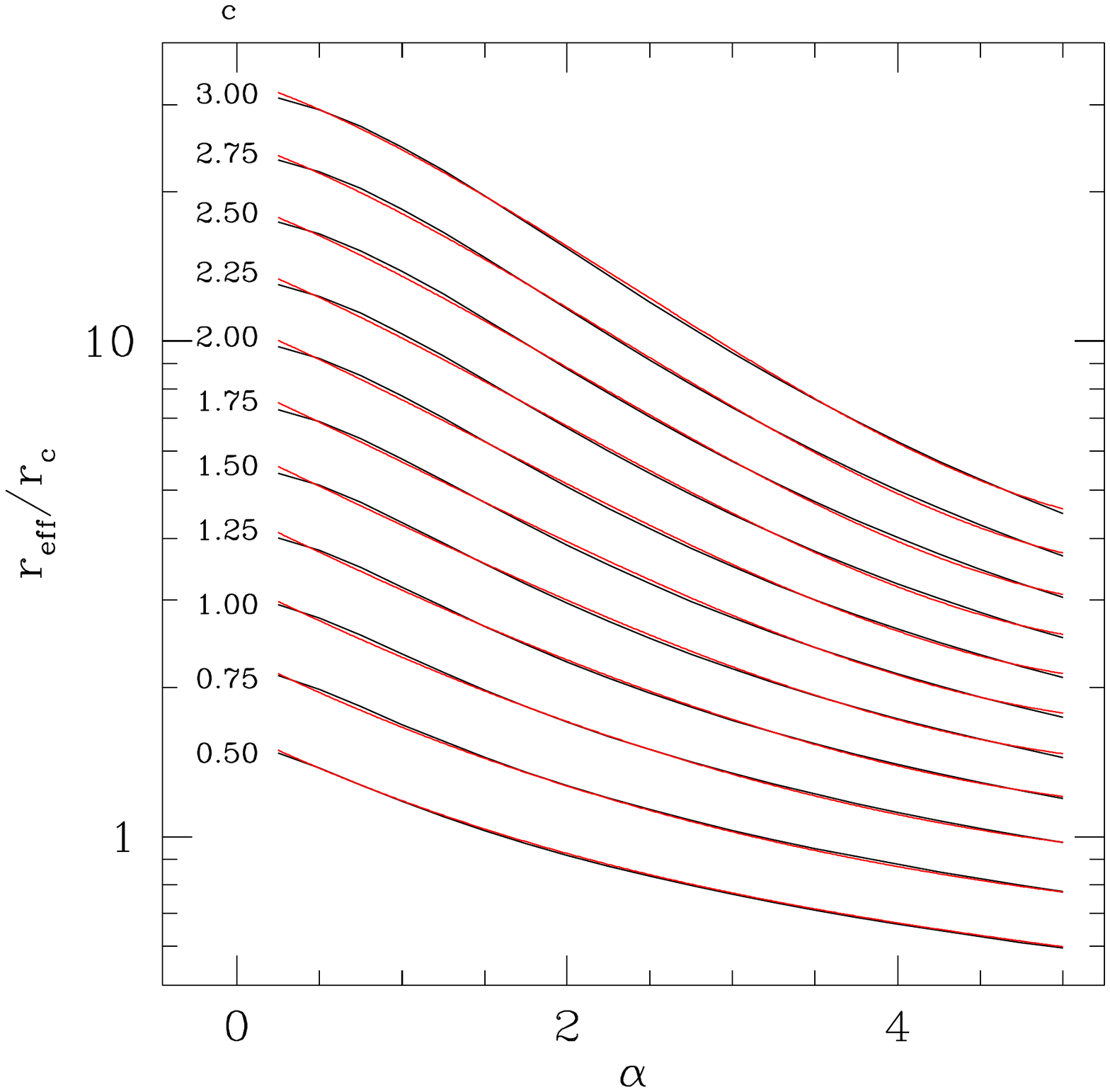}\\
\includegraphics[width=1.04\linewidth,trim=20 40 30 40,clip]{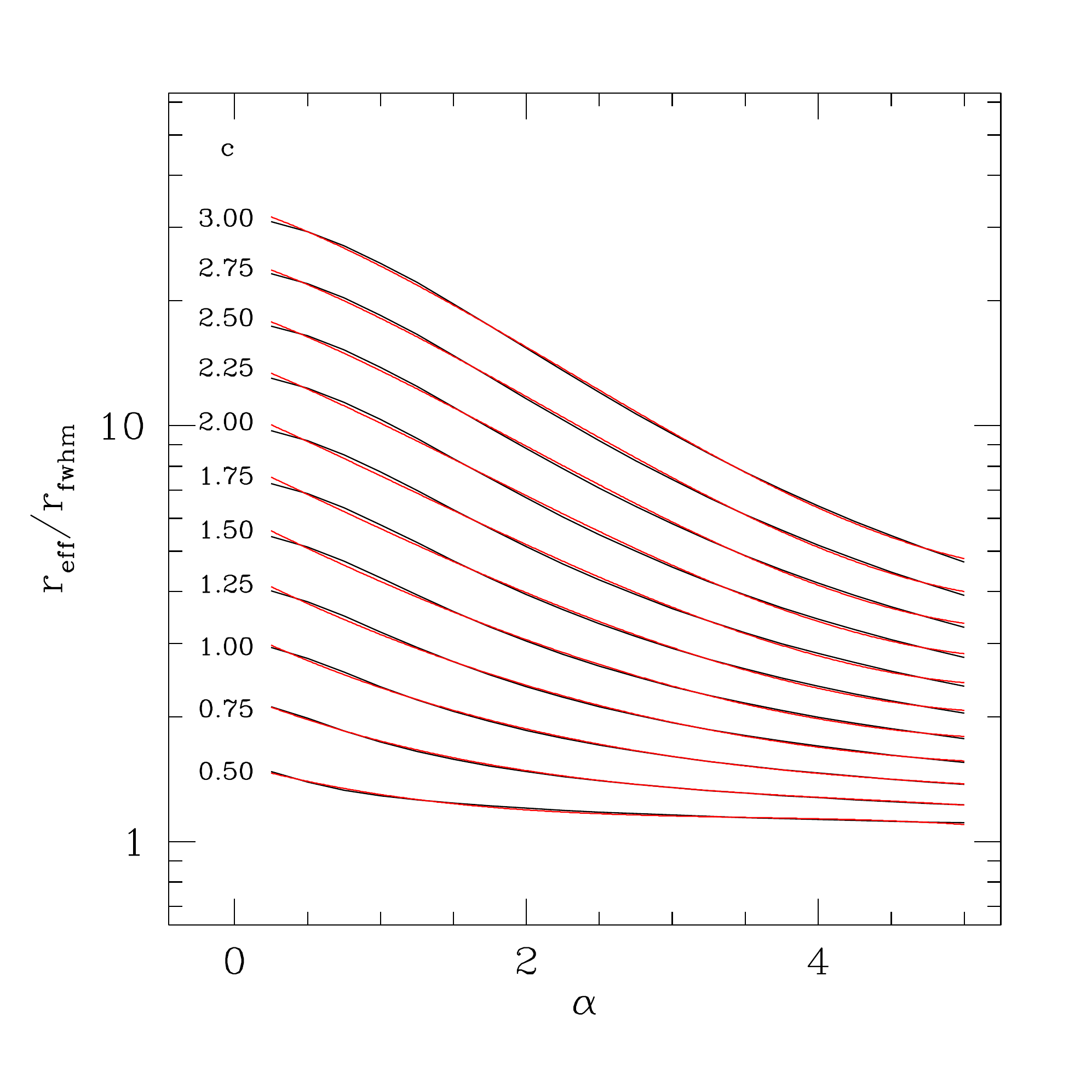}
\caption{Illustration of the relation between $r_{\rm eff}$ and $r_{\rm c}$ (left) 
and between $r_{\rm fwhm}$ and $r_{\rm c}$ (right) for 
\protect\cite{King62} models for few discrete values of the concentration index $c$.
Black curves show the actual values determined by integration of the King profiles
while the red curves are out fit according to eqs.\ref{eq:reffrc} and \ref{eq:reff_King}.
\label{fig:reffrc_c_King}
}
\end{figure}
%\begin{figure}
%\includegraphics[width=1.02\linewidth,trim=20 40 30 40,clip]{reff.pdf}
%\caption{Caption is the same as in Figure\,\ref{fig:reffrc_c_King}, but the red curves 
%are out fit according to eq.\ref{eq:reff_King}.
%\label{fig:rfwhm_c_King}
%}
%\end{figure}

For the original \cite{King62} model there is no simple analytical connection between the 
effective radius $r_{\rm eff}$ containing half the mass/light in projection and the parameters 
of the King profile (core radius, $r_c$, concentration, $C\equiv\!c=\log_{10}(r_t/r_c)$  
and $\alpha$). Here, we derive this connection through direct integration of \cite{King62} 
profiles and fitting the resulting values  by a polynomial fit. The coefficients of the 
polynomials have been determined through $\chi^2$ minimization. Our fit between 
$r_{\rm eff},\ r_c,\ c,\ \alpha$ was derived for values of $c$ between $0.5<c<3.0$ and $\alpha$ 
between $0.25<\alpha<5.0$. This resulted in the following fit:
\begin{equation}
\log_{10}(r_{\rm eff}/r_c) = c_1+c_2\times\alpha+c_3\times\alpha^2+c_4\times\alpha^3,
\label{eq:reff_King}
\end{equation}
where the coefficients $c_1,\ c_2,\ c_3,\ c_4$ are given by:
\begin{gather*}
\begin{bmatrix}
 c_1 \\ c_2\\ c_3\\ c_4\\
  \end{bmatrix}
 =
 \begin{bmatrix}
-0.13150 &  0.74000 & -0.10900 &  0.01569 \\
-0.13980 & -0.05897 &  0.03980 & -0.00558 \\
  0.02390 &  0.00112 & -0.01689 &  0.00307 \\
-0.00198 &  0.00080 &  0.00166 & -0.00035
 \end{bmatrix}
 \times
 \begin{bmatrix}
1 \\ c \\ c^2 \\ c^3\\
 \end{bmatrix}
\end{gather*}
We used this relation to calculate the nuclear star cluster $r_{\rm eff}$ given in 
Table\,\ref{Table:bestfits}. These relations are good to within a percent. Often, 
fitting can report the measured FWHM, therefore, similarly, we also obtained the 
relation between the effective radius and the FWHM as:
\begin{equation}
\log_{10}(r_{\rm eff}/r_{\rm FWHM})=c_1+c_2\times\alpha+c_3\times\alpha^2+c_4\times\alpha^3,
\label{eq:reffrc}
\end{equation}
with the following values for the coefficients $c_1,\ c_2,\ c_3,\ c_4$:
\begin{gather*}
\begin{bmatrix}
 c_1 \\ c_2\\ c_3\\ c_4\\
  \end{bmatrix}
 =
 \begin{bmatrix}
-0.20070 &  0.85499 & -0.16590 &  0.02455 \\
  0.05510 & -0.40780 &  0.22090 & -0.03462 \\
  0.00264 &  0.07583 & -0.06395 &  0.01131  \\
-0.00143 & -0.00456 &  0.00559 & -0.00109
 \end{bmatrix}
 \times
 \begin{bmatrix}
1 \\ c \\ c^2 \\ c^3.
 \end{bmatrix}
\end{gather*}
An illustration of these relations we show in Figure\,\ref{fig:reffrc_c_King} for few 
fixed concentration indexes. As it can be seen from the figure, the achieved precision 
is much less than 1\%.

%%%%%%%%%%%%%%%%%%%%%%%%%%%%%%%%%%%%%%%%%%%%%%%%%%

% Don't change these lines
\bsp	% typesetting comment
\label{lastpage}
\end{document}